\documentclass[twocolumns]{aa}
\usepackage{graphicx}
\usepackage{txfonts}
\begin{document}

\title{The spatial clustering of mid-IR selected star forming galaxies at $z\sim1$ in the GOODS fields}

\author{R. Gilli\inst{1}, E. Daddi\inst{2}, R. Chary\inst{3},
M. Dickinson\inst{4}, D. Elbaz\inst{2}, M. Giavalisco\inst{5},
M. Kitzbichler\inst{6}, D. Stern\inst{7},
E. Vanzella\inst{8} 
}

\authorrunning{R. Gilli et al.}  
\titlerunning{Spatial clustering of $z\sim1$ mid-IR galaxies}

   \offprints{R. Gilli \\email:{\tt roberto.gilli@oabo.inaf.it}}

   \date{Received ... ; accepted ...}

\institute{
Istituto Nazionale di Astrofisica (INAF) - Osservatorio
Astronomico di Bologna, Via Ranzani 1, 40127 Bologna, Italy 
\and
Laboratoire AIM, CEA/DSM - CNRS - Universit\'e Paris Diderot,
DAPNIA/SAp, Orme des Merisiers, 91191 Gif-sur-Yvette, France 
\and 
Spitzer Science Center, California
Institute of Technology, Mail Stop 220-6, Pasadena, CA 91125, USA 
\and
National Optical Astronomy Observatory, P.O. Box 26732, Tucson, AZ
85726, USA 
\and 
University of Massachusetts, Astronomy Dept, Amherst, MA 01003, USA
\and 
Max-Planck Institut f\"ur Astrophysik,
Karl-Schwarzschild-Strasse 1, D-85748 Garching b. M\"unchen, Germany
\and 
Jet Propulsion Laboratory, California
Institute of Technology, Pasadena, CA 91109, USA 
\and 
Istituto
Nazionale di Astrofisica (INAF) - Osservatorio Astronomico di Trieste,
Via G. Tiepolo 11, 34131 Trieste, Italy 
}

\abstract{We present the first spatial clustering measurements of
$z\sim1$, 24\,$\mu$m-selected, star forming galaxies in the Great
Observatories Origins Deep Survey (GOODS). The sample under
investigation includes 495 objects in GOODS-South and 811 objects in
GOODS-North selected down to flux densities of $f_{24}>20\,\mu$Jy and
$z_{AB}<23.5$\,mag, for which spectroscopic redshifts are available.
The median redshift, IR luminosity and star formation rate (SFR) of
the samples are $z\sim 0.8$, $L_{IR}\sim4.4\times10^{10}$ $L_{\odot}$,
and SFR$\sim 7.6\,M_{\odot}$ yr$^{-1}$, respectively. We measure the
projected correlation function $w(r_p)$ on scales of
$r_p=0.06-10\,h^{-1}$ Mpc, from which we derive a best fit comoving
correlation length of $r_0 = 4.0\pm0.4\,h^{-1}$ Mpc and slope of
$\gamma=1.5\pm0.1$ for the whole $f_{24}>20\,\mu$Jy sample after
combining the two fields. We find indications of a larger correlation
length for objects of higher luminosity, with Luminous Infrared
Galaxies (LIRGs, $L_{IR}>10^{11}\,L_{\odot}$) reaching $r_0\sim
5.1\,h^{-1}$ Mpc. This would imply that galaxies with larger SFRs are
hosted in progressively more massive halos, reaching minimum halo
masses of $\sim3\times10^{12}M_\odot$ for LIRGs. We compare our
measurements with the predictions from semi-analytic models based on
the Millennium simulation. The variance in the models is used to
estimate the errors in our GOODS clustering measurements, which are
dominated by cosmic variance. The measurements from the two GOODS
fields are found to be consistent within the errors. On scales of the
GOODS fields, the real sources appear more strongly clustered than
objects in the Millennium-simulation based catalogs, if the selection
function is applied consistently. This suggests that star formation at
$z\sim0.5$--1 is being hosted in more massive halos and denser
environments than currently predicted by galaxy formation
models. Mid-IR selected sources appear also to be more strongly
clustered than optically selected ones at similar redshifts in deep
surveys like the DEEP2 Galaxy Redshift Survey and the VIMOS-VLT Deep
Survey (VVDS), although the significance of this result is $\lesssim
3\sigma$ when accounting for cosmic variance. We find that LIRGs at
$z\sim1$ are consistent with being the direct descendants of Lyman
Break Galaxies and UV-selected galaxies at $z\sim2$--3, both in term
of number densities and clustering properties, which would suggest
long lasting star-formation activity in galaxies over cosmological
timescales. The local descendants of $z\sim0.5$--1 star forming
galaxies are not luminous IR galaxies but are more likely to be
normal, $L<L_*$ ellipticals and bright spirals.


}

   \maketitle

\section{Introduction}\label{introduction}

In the general paradigm of large scale structure formation, the small
primordial fluctuations in the matter density field progressively grow
through gravitational collapse, leading to the present-day complex
network of clumps and filaments which is often referred to as the
``Cosmic Web''. Baryons are believed to cool within dark matter halos
(DMHs) and form galaxies and cluster of galaxies, whose distribution
on the sky should then trace that of the underlying dark matter. While
the formation and the evolution of dark matter structures can be
followed in a relatively straightforward way through N-body simulations
(e.g., Jenkins et al.\ 1998; Springel et al.\ 2005), which can be also
approximated analytically with high accuracy (Peacock \& Dodds 1996),
the physics of baryon cooling and galaxy formation within DMHs is far
more complex. As a result of these complex physical processes, the
distribution of galaxies in the sky may be biased with respect to that
of the underlying matter distribution. The amplitude of this bias is
expected to evolve with cosmic time and be dependent on galaxy type,
luminosity and local environment (Norberg et al.\ 2002). 
The comparison between the clustering properties of
galaxies and those of DMHs predicted by cold dark matter (CDM) models
can be used to evaluate the typical mass of the DMHs in which galaxies
form and reside as a function of cosmic time. Following the evolution
of the typical DMH hosting a given galaxy type at any given time also
allows one to predict the environment in which that galaxy should be
found nowadays and the environment in which it was residing in the
past. In other words, under reasonable assumptions, it is possible to
guess the progenitors and descendants of galaxy populations
observed at any cosmological epoch.

Galaxy clustering has been traditionally studied by means of the
two-point correlation function $\xi(r)$, defined as the excess
probability over random of finding a pair of galaxies at a separation
$r$ from one another, which is often approximated with a power law of
the form $\xi(r)=(r/r_0)^{-\gamma}$. In the local Universe different
clustering properties have been observed as a function of galaxy
type. In the Sloan Digital Sky Survey (SDSS, York et al.\ 2000), at a
median redshift of $z\sim0.1$, red, early-type galaxies show a larger
correlation length and a steeper slope ($r_0=6.8\,h^{-1}{\rm Mpc
,}\:\gamma=1.9$) than blue, late type galaxies ($r_0=4.0\,h^{-1}{\rm
Mpc ,}\:\gamma=1.4$; Zehavi et al.\ 2002). Similar results arise from
the 2dF Galaxy Redshift Survey (2dFGRS; Colless et al.\ 2001), in
which, at a similar median redshift, passive galaxies show a
correlation length and slope of $r_0=6.0\,h^{-1}{\rm Mpc
,}\:\gamma=1.9$ as opposed to $r_0=4.1\,h^{-1}\rm{Mpc ,}\:\gamma=1.5$
measured for star forming galaxies (Madgwick et al.\ 2003).

At cosmologically significant distances, deep surveys on sky patches
of less than 1~deg$^2$, complemented by large spectroscopic campaigns,
are measuring the clustering of high redshift objects with reasonable
accuracy.  The separation between the clustering properties of star
forming and passively evolving galaxies seems to be well established
even at redshifts around 1. In the DEEP2 Galaxy Redshift Survey, Coil
et al.\ (2004) found that $z\sim 0.9$ galaxies with absorption line
spectra have a correlation length significantly larger than
emission-line galaxies at the same redshift. A similar result has been
found in the VIMOS-VLT Deep Survey (VVDS) by Meneux et al.\ (2006),
who measured a correlation length that was larger for red galaxies
than for blue galaxies at $z\sim 0.8$.

Porciani \& Giavalisco (2002) and Adelberger et al.\ (2005) measured
the clustering properties of star forming galaxies selected by the
Lyman-break technique between redshifts of 1.7 and 3 (see also Hamana
et al.\ 2004, Ouchi et al.\ 2005 and Lee et al.\ 2006 for Lyman break
galaxies selected at $z\sim4-5$). The measured comoving correlation
length of 4.0-4.5 $h^{-1}$ Mpc for these high redshift objects is
expected to increase with time and suggests that they will evolve into
moderate-luminosity, elliptical galaxies by $z=0$ (Adelberger et al.\
2005).

While all of the above described samples are based on optical
selection, star formation in galaxies can be efficiently traced by
mid-infrared observations. The star formation rate (SFR), particularly
the dust obscured component, is indeed directly correlated to the
mid-IR luminosity, which is in turn a robust proxy for the total
(8-1000$\mu$m) IR luminosity (e.g., Spinoglio et al.\ 1995; Chary \&
Elbaz 2001; Forster-Schreiber et al.\ 2004). This has been
demonstrated in the present-day Universe, but seems to hold at least
up to $z\sim 1$, where the bulk of star-formation occurs in
dust-obscured regions. Indeed, the deepest existing radio data have
shown that $L_{IR}$ values determined from the mid-IR luminosity of
galaxies at $z\sim1$ are consistent with those derived using the radio
to IR luminosity correlation (Elbaz et al.\ 2002, Appleton et al.\
2004).

Early work by the Infrared Astronomical Satellite (IRAS) showed that
the correlation length of nearby (median $z\sim0.03$) mid-IR bright
galaxies ($f_{60\mu m}>1.2$ Jy) is about 4 $h^{-1}$ Mpc (Fisher et
al.\ 1994), in agreement with the values measured for local star
forming objects by the SDSS and 2dFGRS. More recently, an attempt to
measure the clustering properties of mid-IR selected sources at
fainter flux densities has been made (D'Elia et al.\ 2005). Based on a
small sample of galaxies detected by the Infrared Space Observatory
(ISO) with $f_{15\mu m}>0.5$ mJy, D'Elia et al. found that the
clustering level measured for these $z\sim 0.2$ galaxies is similar to
that measured by IRAS for more local sources.

The {\it Spitzer Space Telescope} (Werner et al. 2004), with its
unprecedented sensitivity at mid-IR and far-IR wavelengths, is
enabling further progress to be made. Deep surveys at 24$\mu$m are
being carried out in different regions of the sky (see, e.g., Papovich
et al.\ 2004), with the deepest ones being performed in the two GOODS
fields down to $f_{24}\sim 10-20\mu$Jy (Chary et al.\ in
preparation). For the first time, this allows us to select field
galaxies based on their ongoing level of star formation activity at a
wavelength where dust corrections will be negligible. This is a more
physically motivated selection than those based on qualitative galaxy
properties like color bi-modality. It thus provides greater insight
into the nature of galaxy and star formation in the distant Universe
and a more straightforward comparison to galaxy formation models. Our
goal is to investigate the spatial distribution of $z\lesssim 1$ star
forming galaxies, and assess the dependence between environment and
star-formation rate.  By constraining the nature of the descendants of
star forming galaxies at $z\lesssim1$, we provide insight into the
nature of downsizing of galaxy formation, a well established pattern
for galaxy evolution which implies that star formation is taking place
preferentially in more massive galaxies at higher redshifts (e.g.,
Cowie et al.\ 1996). A tight correlation between galaxy mass and
star-formation rate has been discovered, with slope close to unity.
This correlation has been shown to exist both in the local Universe as
well as at $z\sim1.2$ (Noeske et al.\ 2007; Elbaz et al.\ 2007) with
tentative evidence that it may be valid even at $z\sim2$ (Daddi et
al.\ 2007a). As more massive galaxies are on average hosted in more
massive halos, we expect to find a direct correlation between
clustering strength and star formation rate in the distant Universe.

Given the large (5-6 arcsec FWHM) resolution of the MIPS instrument
(Multiband Imaging Photometer for {\it Spitzer}; Rieke et al.\ 2004)
confusion problems due to blending are severe at the faintest flux
densities. This makes a proper measure of the angular correlation
function of faint MIPS sources difficult, leaving the 3D correlation
function as the most viable method for estimating their clustering
properties.  In this paper, we measure the spatial clustering of
24$\mu$m selected sources in the two GOODS fields by means of the
projected correlation function $w(r_p)$. Blending problems at short
scales are completely overcome in this case, as angular clustering
terms are negligible as discussed later in the paper.

The paper is organized as follows: in Section~2 we
describe the data sets and the selection criteria adopted to define
the samples used in the clustering analysis. In Section~3 we present
the methods utilized to estimate the correlation function. In
Section~4 several safety checks are performed to validate the adopted
techniques. Simulations are also run to estimate errors on our
measurements due to cosmic variance. The results of our analysis are
presented in Section~5. In Section~6 the clustering measurements are
discussed, interpreted and compared to estimates from optical
surveys. The conclusions are presented in Section~7.

Throughout this paper, a flat cosmology with $\Omega_m=0.3$ and
$\Omega_{\Lambda}=0.7$ is assumed. Unless otherwise stated, we refer
to comoving distances in units of $h^{-1}$ Mpc, where $H_0=100\;h$ km
s$^{-1}$ Mpc$^{-1}$. Luminosities are calculated using $h=0.7$.

\begin{figure}[t]
\includegraphics[width=8.5cm]{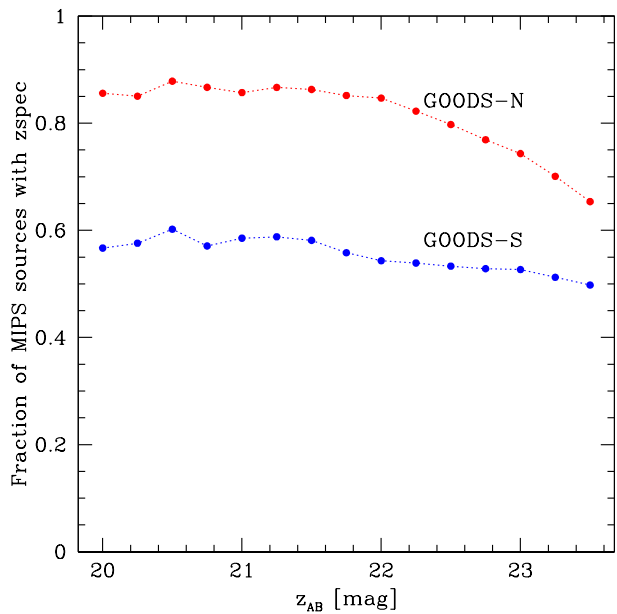}
\caption{Spectroscopic completeness down to $z_{AB}=23.5$\,mag for
galaxies with $f_{24}>20\,\mu$Jy in GOODS-N (upper curve) and GOODS-S
(lower curve).}
\label{comp}
\end{figure}

\begin{figure}[t]
\includegraphics[width=9cm]{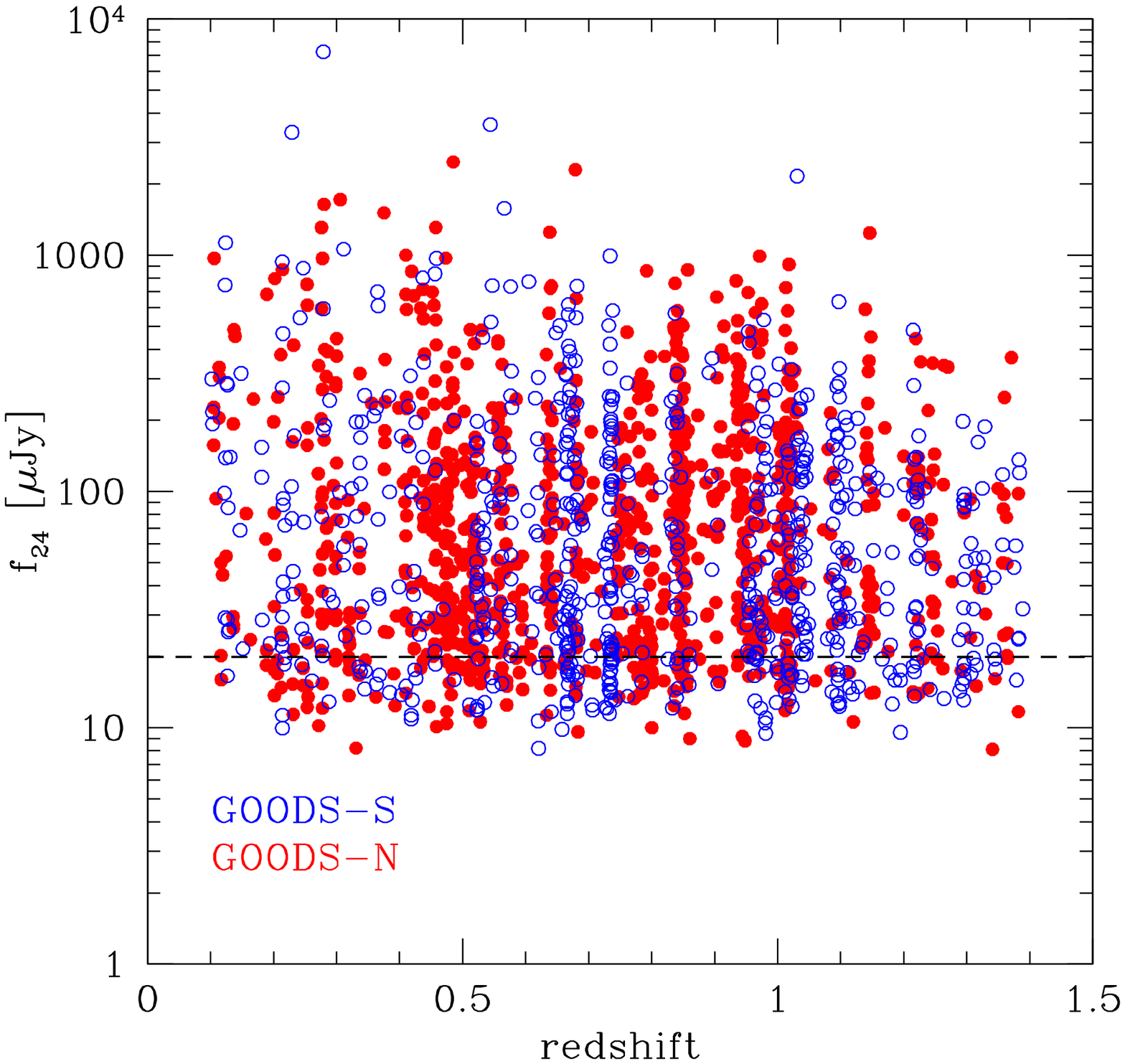}
\caption{24$\mu$m flux density vs redshift for sources detected
by {\it Spitzer}/MIPS in GOODS-S (open circles) and GOODS-N (filled
circles). Only sources with spectroscopic redshifts are shown. The
dashed line shows the $f_{24}=20\,\mu$Jy flux limit used in this
work.}
\label{zf}
\end{figure}

\begin{figure}[t]
\includegraphics[width=9cm]{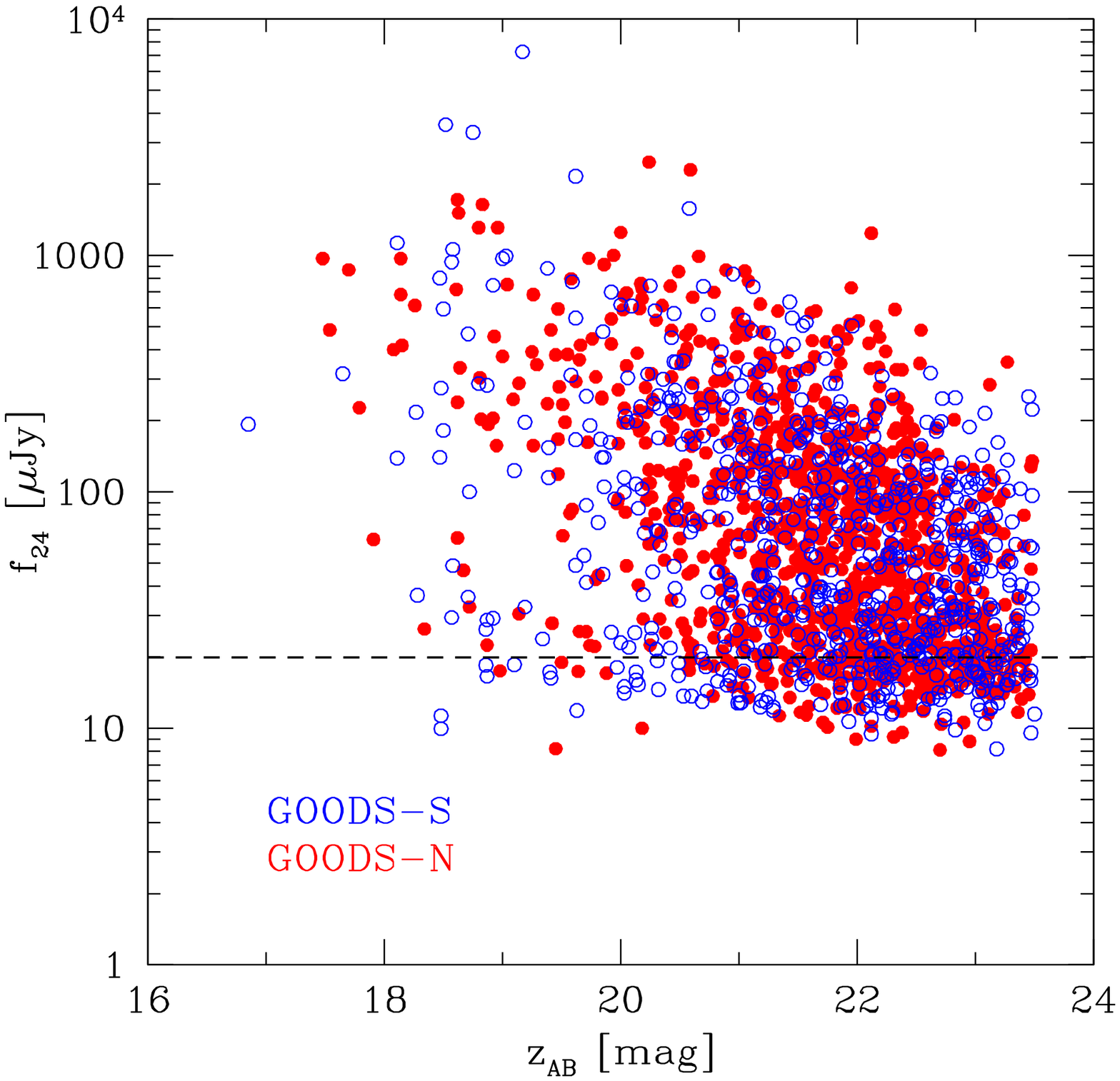}
\caption{24$\mu$m flux density vs $z_{AB}$ magnitude for sources
detected by {\it Spitzer}/MIPS in GOODS-S (open circles) and GOODS-N
(filled circles). Only sources with spectroscopic redshifts are
shown. The dashed line shows the $f_{24}=20\,\mu$Jy flux limit
used in this work.}
\label{magzf}
\end{figure}

\begin{figure}[t]
\includegraphics[width=9cm]{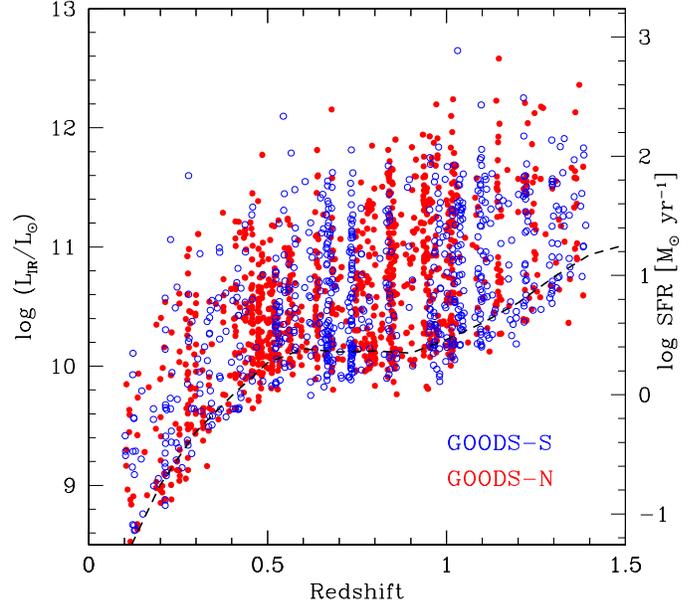}
\caption{$L_{IR}$ vs redshift for sources detected by
{\it Spitzer}/MIPS in GOODS-S (open circles) and GOODS-N (filled
circles). Only sources with spectroscopic redshifts are shown. The
dashed line shows the $f_{24}=20\,\mu$Jy flux limit used in this
work.}
\label{lirz}
\end{figure}

\section{The samples}\label{samp}

The GOODS-South and GOODS-North fields, each covering about
$10\times16$ arcmin, have been observed by {\it Spitzer} as part of
the Great Observatories Origins Deep Survey Legacy Program (Dickinson
et al.\ 2007, in preparation). Deep 24 $\mu$m observations with MIPS
were carried out down to sensitivities of $\sim 12\,\mu$Jy
($\sim3\sigma$) in both fields (Chary et al., in preparation). Source
catalogs at shorter wavelengths (Dickinson et al., in preparation)
based on the Infrared Array Camera (IRAC; Fazio et al.\ 2004), were
used as prior positions in order to improve source deblending and
identify unique counterparts. Spectroscopic redshifts have been
collected for about 60\% of the MIPS sources with $z_{AB}<23.5$\,mag
from a compilation of all the different follow-up spectroscopy
programs carried out in the GOODS fields. In particular, for the
GOODS-S field, we use the spectroscopic redshifts made available by Le
Fevre et al.\ (2004), Mignoli et al.\ (2005), Vanzella et al.\ (2005;
2006). Redshifts in GOODS-N have been published in many papers over
the years. At the redshifts of interest in this paper, the largest
portion of the published redshifts can be found in Cohen et al.\
(2000), Wirth et al.\ (2004), and Cowie et al.\ (2004). We supplement
these with additional redshifts for 24$\mu$m selected sources from
Stern et al.\ (in preparation). The spectroscopic completeness down to
$z_{AB}=23.5$ mag is shown in Fig.~\ref{comp}. In both fields the
completeness level decreases towards fainter magnitudes, but in
GOODS-N it is systematically higher than than in GOODS-S. For sources
with $z_{AB}<23.5$\,mag the completeness level in GOODS-N is 65\%,
compared to 50\% in GOODS-S. Only sources at $0.1<z<1.4$ were
considered in this work. The $z<1.4$ limit is imposed in order to
remain in a redshift range where the spectroscopic sampling is
highest, and where the observed 24$\mu$m flux density can be used as
an accurate tracer of the total IR luminosity of galaxies.  Although
$24\mu$m observations can be used to obtain reasonable measurements of
star formation activity averaged over the galaxy population at even
higher redshifts (e.g., Daddi et al 2005), individual sources with
anomalous properties may show significant errors in their derived
$L_{IR}$ (Daddi et al 2007a, Papovich et al.\ 2007).  Redshift quality
flag information is available for most of the spectroscopic surveys
done in GOODS-S, but is missing for some of the surveys in GOODS-N. In
GOODS-S we considered only objects with high quality flags. In GOODS-N
we have excluded some galaxies ($<1$\% of the total sample) which
appear to have incorrect spectroscopic redshifts, based on the shape
of their spectral energy distribution and photometric
redshifts. Furthermore, we have limited our analysis to sources with
$f_{24}>20\mu$Jy, for which the flux density estimate is reliable and
source confusion is well understood (Chary 2006). About 20\% of the
sources fall below this limit and are therefore excluded from our
clustering analysis. In total, 558 objects in GOODS-S and 875 objects
in GOODS-N are found to satisfy these selection criteria (including
AGN, see later).  After accounting for spectroscopic incompleteness,
the number of $f_{24}>20\mu$Jy sources in GOODS-S and GOODS-N differ
by $\sim 20\%$. As shown in Section~\ref{varsec}, this is consistent
with being due to cosmic variance.

\begin{figure*}[t]
\includegraphics[width=9cm]{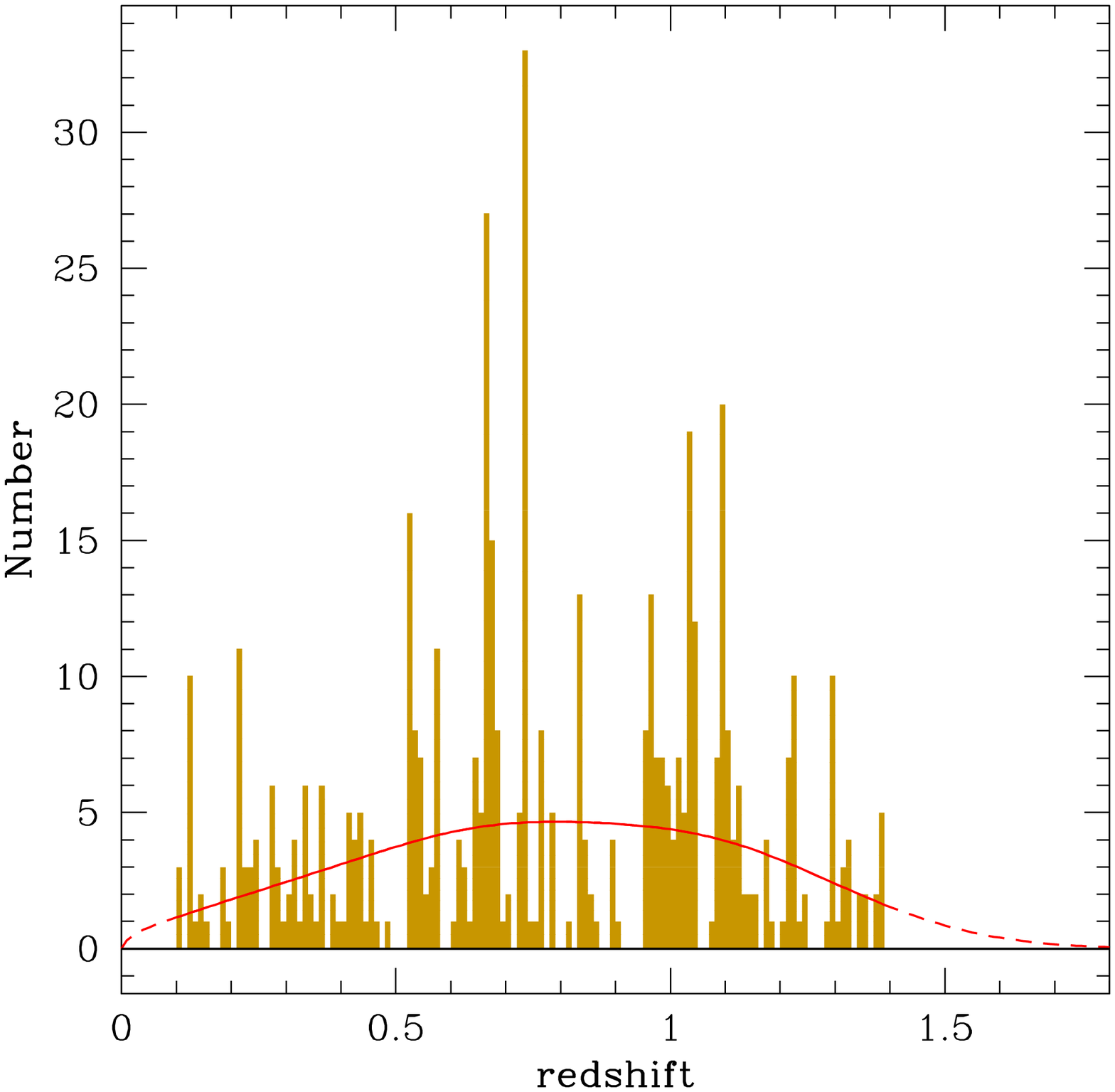}
\includegraphics[width=9cm]{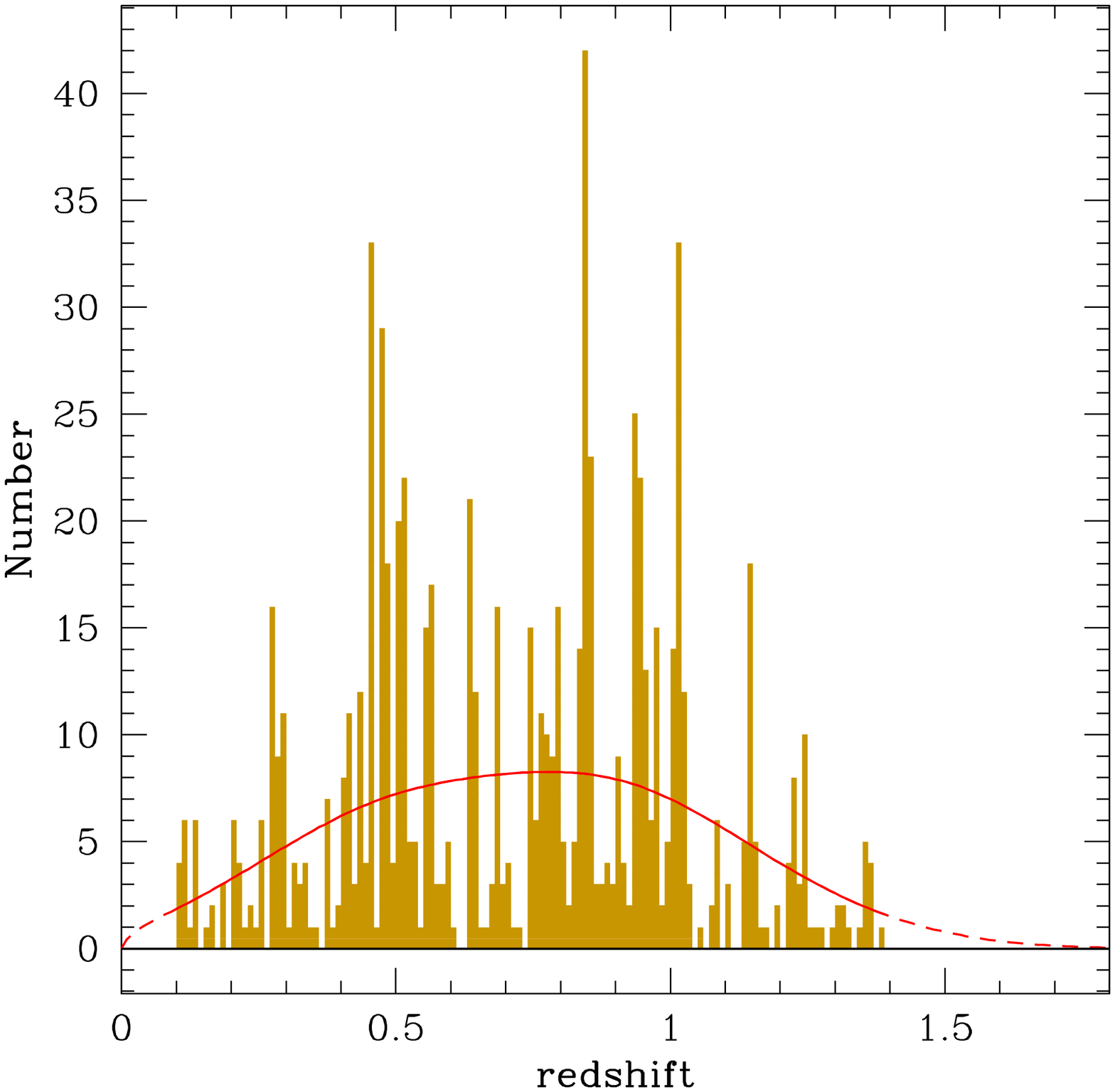}
\caption{Redshift distribution for MIPS sources with $f_{24}>20\mu$Jy
and spectroscopic redshift in the GOODS-S ({\it left}) and GOODS-N
({\it right}) fields (AGN excluded), binned to $\Delta z = 0.01$. The
smooth curves show the selection function adopted to generate the
random control sample, obtained by smoothing the observed redshift
distributions and truncated at $z<0.1$ and $z>1.4$ as the data
samples.}
\label{zdist}
\end{figure*}

In Fig.~\ref{zf} and Fig.~\ref{magzf} the 24$\mu$m flux densities of
sources in the two GOODS fields are plotted against their spectroscopic
redshifts and $z_{AB}$ magnitudes, respectively. Fainter 24$\mu$m
sources have on average fainter optical counterparts and tend to be
at higher redshifts, although the redshift dependence of the
average 24$\mu$m flux density appears rather weak. Several redshift
structures can be immediately identified, which are also traced by
sources selected at other wavelengths (e.g., Cohen et al.\ 1996;
Gilli et al.\ 2003; Barger et al.\ 2003). The 24$\mu$m flux density and
redshift distribution in the two fields are similar (see also
Fig.~\ref{zdist}). The median 24$\mu$m flux density, optical magnitude
and redshift for the considered samples are $f_{24}\sim$ 74 $\mu$Jy,
$z_{AB}\sim$21.8\,mag and $z\sim0.75$, respectively.  We 
compute the total (8-1000$\mu$m) IR luminosity
$L_{IR}$ from the observed 24\,$\mu$m flux density, assuming the 
luminosity-dependent model templates of Chary \& Elbaz (2001). 
The total IR luminosity 
provides a measure of the star formation rate in the galaxy
using the relation SFR$=L_{IR}\times 1.72\times10^{-10}\,M_{\odot}$~yr$^{-1}$
(Kennicutt et al.\ 1998). We note that if more recent estimates of the
stellar initial mass function are adopted (Kroupa 2001, Chabrier
2003), the same $L_{IR}$ systematically converts into a $\sim 30$\%
lower SFR.
The exact conversion rate does not have an important effect on our
results.  The $L_{IR}$ (SFR) versus redshift plot for the galaxy
sample considered here
is shown in Fig.~\ref{lirz}, along with the
$L_{IR}$ cut introduced at each redshift by the $f_{24}>20\mu$Jy
selection.  The luminosity distribution is similar in the two
fields. The median luminosity and star formation rate are
$4.4\times10^{10}\;L_{\odot}$ and 7.6 $M_{\odot}$ yr$^{-1}$, respectively.
About 90\% of the objects in the two fields have
$L_{IR}>10^{10}\;L_{\odot}$ while about 30\% have
$L_{IR}>10^{11}\;L_{\odot}$. The latter are classified as Luminous Infrared
Galaxies (LIRGs), and are forming stars at an average
 rate of $\sim$35 $M_{\odot}$ yr$^{-1}$.

We note that the SFR estimated from the $L_{IR}$ values may
be a lower limit to the true galaxy SFR since 
it excludes the unobscured star-formation traced by the observed
UV emission. We therefore considered B-band
data from the Advanced Camera for Surveys (ACS) onboard the Hubble
Space Telescope (HST), which traces the rest frame UV flux for galaxies
at $z>0.5$, i.e., for the majority of the sources in our sample. We
found that the SFR increases by only 4\% on average when including
the ACS data. We also note that the fraction of galaxies for which the
SFR may have been underestimated significantly (e.g., by a factor of
1.5-2), is less that 4\%. 
Due to the fact that the UV flux may have a contribution from
old, evolved stars, these correction factors are upper limits. 
Our estimates appear to be
in good agreement with those of Bell et al.\ (2005), who derive an
average UV contribution of 5-10\% to the global (mid-IR + UV) SFR of
$z\sim 0.7$ star forming galaxies observed by {\it Spitzer}. Furthermore,
since the UV correction decreases with increasing SFR, it becomes
completely negligible for LIRGs. To summarize, UV corrections to the
SFR do not have a significant impact on our results and are
therefore neglected in the following analysis.

While most of these mid-IR selected sources are expected to be star
forming galaxies (elliptical galaxies should be virtually absent from
mid-IR selected samples), a significant fraction of sources may be
active galactic nuclei (AGN), in which the radiation absorbed by
circumnuclear material is re-emitted in the IR regime. Based on the
X-ray properties of sources, we therefore tried to eliminate AGN
interlopers. Both fields have been observed by {\it Chandra} with
extremely deep (1-2 Msec) exposures (Giacconi et al.\ 2002, Alexander
et al.\ 2003). Using an AGN classification similar to that adopted in
Gilli et al.\ (2005), we flagged as AGN those sources with either
observed 0.5-10 keV luminosities above $10^{42}$ erg s$^{-1}$ or with
a column density above $N_H=10^{22}$ cm$^{-2}$. The column density was
estimated by assuming an intrinsic AGN template with spectral index of
0.7 and absorbing it at the source redshift to reproduce the observed
hard-to-soft X-ray flux ratio.  About 8\% of the sources were removed
from the samples using this AGN classification. We nonetheless
verified that, due to the small fraction of AGN candidates, our
results are insensitive to the methodology adopted to remove
AGN. Moreover, our conclusions do not vary significant even if AGN are
not excluded from the sample.

After the AGN are removed, we are left with samples of 495 and 811
galaxies, in GOODS-South and North, respectively. One may wonder if
our samples are significantly contaminated by AGN which went
undetected in the X-rays. Indeed, Alonso-Herrero et al.\ (2006) in
GOODS-S and Donley et al.\ (2007) in GOODS-N, respectively, have
identified a large population of IR luminous galaxies showing
power-law emission in the IRAC $3.6-8\mu$m bands.  The power-law
emission is thought to be due to hot dust in the vicinity of the
AGN. Yet, half of these sources do not have an X-ray counterpart. We
verified that none of these power-law AGN are present in our
samples. We note that the Donley et al.\ (2007) and Alonso-Herrero et
al.\ (2007) samples are based on shallow $24\mu$m data, span a broader
redshift range and primarily include objects with photometric
redshifts.  In contrast, our galaxies sample much fainter 24\,$\mu$m
flux densities and have spectroscopic redshifts of $z<1.4$.  We are in
the process of defining IR-based AGN samples in our deep MIPS
catalogs. Preliminary analysis suggests that $\lesssim 10\%$ of
sources might be flagged as additional AGN candidates and in principle
should be removed from our samples. Their impact on the clustering
measurements presented in this work is unlikely to be significant and
will be discussed elsewhere when the AGN catalogs are finalized.  Very
recently, Daddi et al.\ (2007b) have shown that a population of highly
obscured AGN, which are both undetected in the X-rays and do not show
a power-law continuum in the IRAC bands, hide in about 20-30\% of IR
luminous ($L_{IR}\gtrsim10^{12}\;L_{\odot}$) galaxies at $z\sim 2$,
providing a significant contribution to their 24$\mu$m emission (see
also Fiore et al.\ 2007).
Given the relatively low IR luminosities ($L_{\rm
IR}\sim10^{10-11}L_\odot$) and the longer mid-IR rest-frame
wavelengths probed here at $z\sim0.7$, we expect that the effect of
contamination from
an obscured AGN population will be less important for our
study.

It should also be noted that we are measuring the clustering
properties of mid-IR selected galaxies over a broad redshift range
from $z=0.1$ to $z=1.4$. Star forming galaxies are undergoing
rapid cosmological evolution in luminosity/density over this
redshift range (e.g., Le Floc'h et
al.\ 2005), and the clustering strength is also likely to
evolve. Although most of the clustering signal measured in
this work is due to galaxy pairs at $z\sim 0.7$, our measurements 
could be
returning a value for the clustering strength that is
an average between 0$<z<1.4$. Thus, our analysis is not
identical to that obtained by considering an ideally
large galaxy sample in a narrow redshift interval around $z\sim
0.7$. This caveat should be borne in mind when comparing our results
with those obtained from other surveys.

\section{Analysis techniques}

To eliminate the distortions introduced by peculiar velocities and
redshift errors, which affect the computation of the
source clustering in redshift space, we resort to the projected
correlation function, defined as in Davis \& Peebles (1983):

\begin{equation}
w(r_p) = \int_{-r_{v0}}^{r_{v0}} \xi(r_p, r_v)dr_v,
\end{equation}
\noindent
where $\xi(r_p, r_v)$ is the two-point correlation function expressed
in terms of the separations perpendicular ($r_p$) and parallel ($r_v$)
to the line of sight, in comoving coordinates.

If the real space correlation function can be approximated by a
power law of the form $\xi(r)=(r/r_0)^{-\gamma}$ and $r_{v0}= \infty$ then
the following relation holds (Peebles 1980):

\begin{equation}
w(r_p) =A(\gamma) r_0^{\gamma} r_p^{1-\gamma},
\end{equation}
\noindent
where $A(\gamma)=\Gamma(1/2)\Gamma[(\gamma-1)/2]/\Gamma(\gamma/2)$ and
$\Gamma(x)$ is Euler's Gamma function. $A(\gamma)$ increases from
3.68 when $\gamma=1.8$ to 7.96 when $\gamma=1.3$. The integration
limit $r_{v0}$ is fixed to 10 $h^{-1}$ Mpc to maximize the
correlation signal (see the end of this Section).

To estimate the correlation function $\xi(r_p,r_v)$ we used the Landy
\& Szalay (1993) estimator, which has been shown to have a nearly
Poissonian variance and which appears to
outperform other popular estimators (e.g.,
see Kerscher et al.\ 2000):
\begin{equation}
\xi(r_p, r_v)=\frac{[DD]-2[DR]+[RR]}{[RR]}.
\end{equation}
\noindent
[DD], [DR] and [RR] are the normalized data-data, data-random and
random-random pairs, i.e.,

\begin{equation}
[DD]\equiv DD(r_p, r_v)\frac{n_r(n_r-1)}{n_d(n_d-1)}
\end{equation}

\begin{equation}
[DR]\equiv DR(r_p, r_v)\frac{(n_r-1)}{2 n_d}
\end{equation}

\begin{equation}
[RR]\equiv RR(r_p, r_v),
\end{equation}
\noindent
while $DD$, $DR$ and $RR$ are the number of data-data, data-random and
random-random pairs at separations $r_p \pm \Delta r_p$ and $r_v \pm
\Delta r_v$; $n_d$ and $n_r$ are the total number of sources in
the data and random sample, respectively.

In order to avoid confusion, we specify how galaxy pairs are counted. 
The number of DD and RR pairs have been
counted only $once$, i.e., the total number of pairs in the real
and in the random samples are $n_d(n_d-1)/2$ and $n_r(n_r-1)/2$,
respectively. This accounts for the factor of 2 in the denominator of
Eq.~5. This way of counting DD and RR pairs has been adopted by
e.g., Landy \& Szalay (1993), Guzzo et al.\ (1997), Gilli et al.\ (2005),
Meneux et al.\ (2006). Other authors, instead, count DD and RR pairs
$twice$, i.e., the total numbers of pairs in their real and random
samples are $n_d(n_d-1)$ and $n_r(n_r-1)$, respectively, which removes
the above mentioned factor of 2 from their formulae. These latter
definitions have been adopted e.g., by Davis \& Peebles (1983),
Kerscher et al.\ (2000), Zehavi et al.\ (2002), Coil et al.\ (2004). It
can be easily shown that the two formulations lead to the same
$\xi(r)$. A simple way to see this is to replace DD and RR in the
formulae of Zehavi et al.\ (2002) and Coil et al.\ (2004) with 2DD' and
2RR', where DD and RR are the numbers of pairs counted twice, while
DD' and RR' are the numbers of pairs counted once (``independent"
pairs).

We note that in Eqs. 4 and 5, $n_d$ is the number of sources
observed in each GOODS field separately. Ideally, instead of using
the observed number of sources, which may produce an overestimate
(underestimate) of the clustering amplitude in under-dense
(over-dense) regions,
one should use 
the true mean source number, which is unknown. 
In principle, averaging the densities of the
GOODS-N and GOODS-S fields would give a better approximation to the
mean source density. However, because of the different spectroscopic
completeness in the two GOODS fields, the estimate of the average
density in a given redshift range may be non-trivial. One possibility
is to assume that the total number of sources in the redshift range
considered in this work ($z=0.1-1.4$) is 20\% larger in GOODS-N than
in GOODS-S. This would be comparable to 
the difference observed in the total surface
density of MIPS sources (after accounting for the 65\% and 50\% total
spectroscopic completeness of GOODS-N and GOODS-S,
respectively). However, since the spectroscopic completeness is 
a function of redshift and optical magnitude, and the completeness
curves are different between the two fields (see Fig.~1), this may not
be the case. At any rate, we have verified that, assuming that the
z=0.1-1.4 source density is 20\% larger in GOODS-N than in GOODS-S,
the use of an averaged density value (i.e., increasing $n_d$ by 10\% in
GOODS-S and decreasing it by the same amount in GOODS-N) gives a
$\sim10\%$ shorter (longer) correlation length in GOODS-S (GOODS-N) than
that estimated by using the density of each field separately. These
fluctuations are of the same order as produced by redshift structures
in our fields (see Section 5.1) and well within the cosmic variance
errors (Section 4). Therefore, they do not change the main
conclusions of the paper.

\begin{figure*}[t]
\includegraphics[width=9cm]{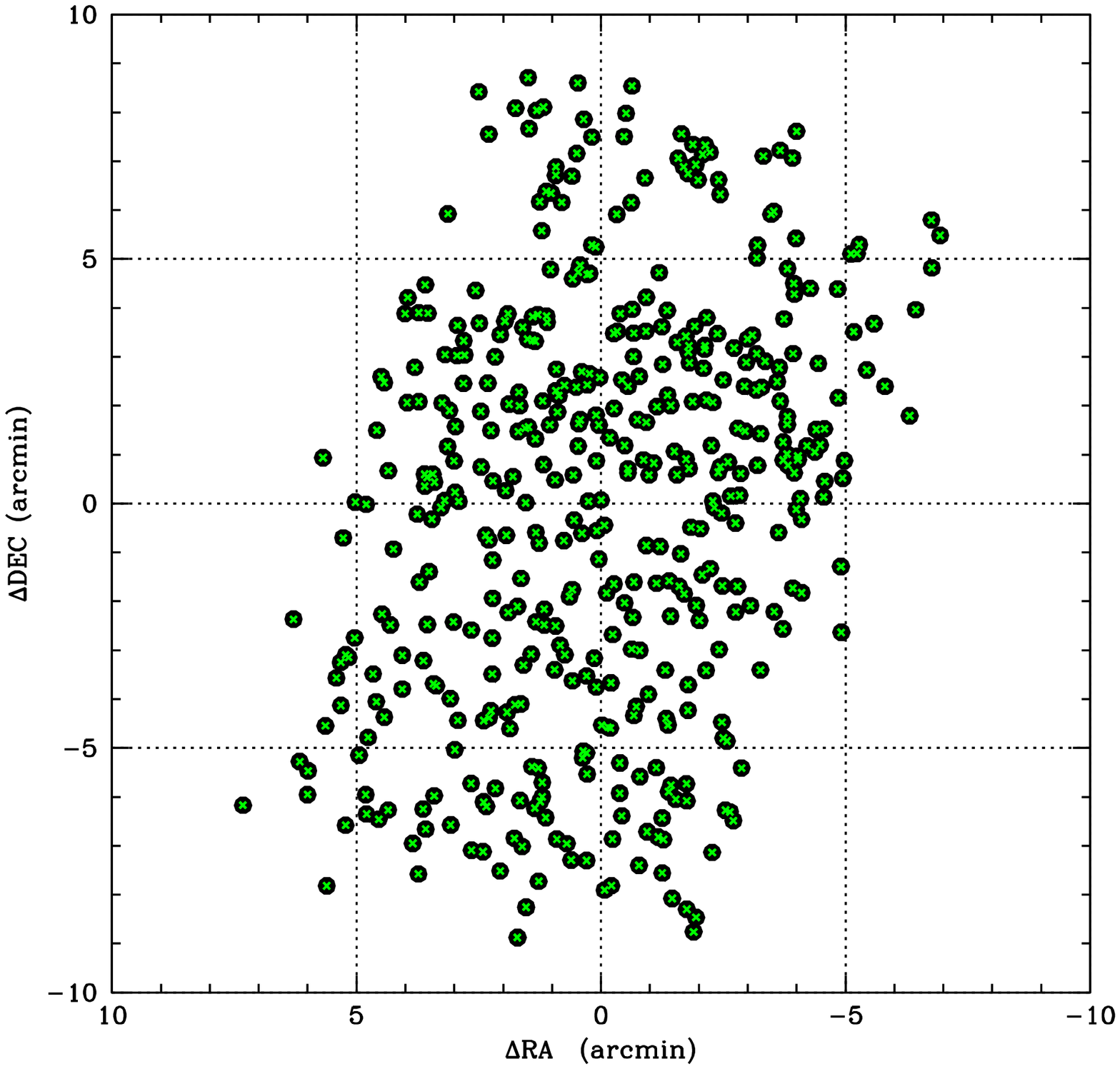}
\includegraphics[width=9cm]{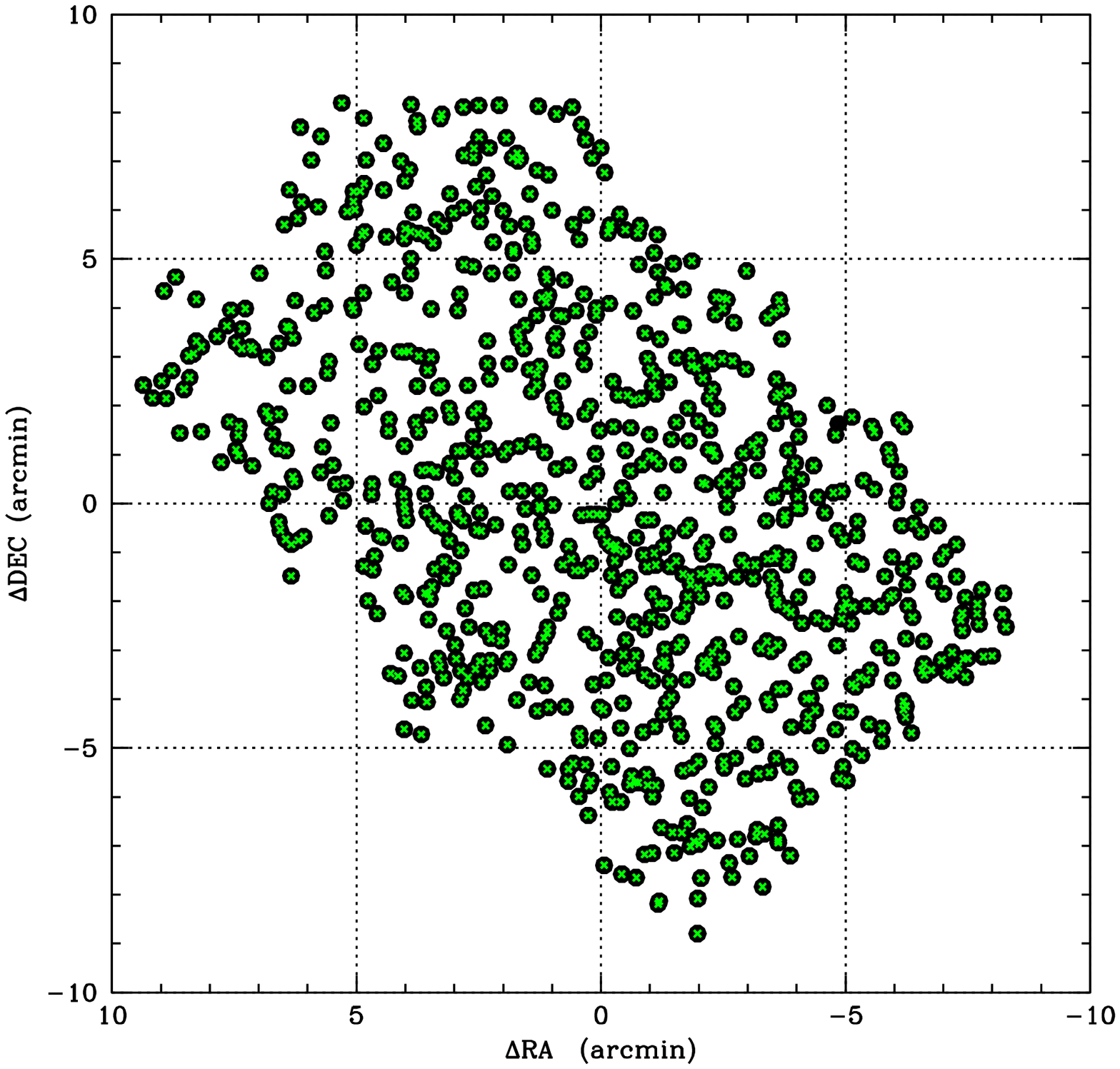}
\caption{Distribution of 24\,$\mu$m sources with
$f_{24}>20\mu$Jy and with spectroscopic redshifts over the GOODS fields. GOODS-S is 
shown in the {\it
left} panel while GOODS-N is shown in the {\it right} panel. The GOODS-S and GOODS-N
fields are centered at ($RA_{J2000},DEC_{J2000}$) =
(53.122368, -27.797262) and (189.215744, 62.234791), 
respectively. Sources in the random samples have been placed at the
coordinates of the real sources.}
\label{skydist}
\end{figure*}

Since both the redshift and the coordinate $(\alpha, \delta)$
distributions of the selected MIPS sources are potentially affected by
observational biases, special care has to be taken in creating the
sample of random sources. We adopted a procedure that has been shown
to work well for X-ray AGN selected in the same fields (see Gilli et
al.\ 2005). The redshifts of the random sources were extracted from a
smoothed distribution of the real one, which should then include the
same observational biases. We assumed a Gaussian smoothing length
$\sigma_z = 0.2$ as a good compromise between smoothing
scales that are too small (which suffer from significant fluctuations due to the observed
redshift spikes) and scales that are too large (where on the contrary the
source density of the smoothed distribution at a given redshift might
not be a good estimate of the average observed value). For each of the
source subsamples considered in this work (see Table 1), we
smoothed the corresponding observed redshift distribution.
The observed and smoothed redshift distributions for the 
$f_{24}>20\mu$Jy samples are shown in Fig.~\ref{zdist}. Due to the
numerous redshift spikes observed, we did not try to measure the
correlation function in different redshift bins since this would be
extremely sensitive to the choice of bin boundaries. The
coordinates ($\alpha,\delta$) of the random sources were extracted
from the coordinate ensemble of the real sample in order to reproduce
the same uneven distribution on the plane of the sky. This procedure
will in principle, reduce the correlation signal, since it removes the
effects of angular clustering. However, as will be verified later,
in deep, pencil-beam surveys like GOODS, where the radial coordinate spans
a much broader distance than the transverse coordinate, most of the
signal is due to redshift clustering, while angular clustering
contributes at most a few percent. The distribution on the sky of
the real sample is shown in Fig.~\ref{skydist}. Each random sample
is built to contain more than 10000 sources.

The source pairs were binned in intervals of $\Delta{\rm
log}\,r_p$=0.1, and $w(r_p)$ was measured in each bin. The resulting data
points were then fit with a power law and the best fit parameters
$\gamma$ and $r_0$ were determined via $\chi^2$ minimization. Given
the small number of pairs which fall into certain bins (especially at the
smallest scales), we used the formulae of Gehrels (1986) to estimate
the 
68\% confidence interval (i.e., $1\sigma$ errorbars in Gaussian statistics).

It is well known that Poisson error bars underestimate the
uncertainties in the correlation function when source pairs are not
independent, which is the case for our sample. More importantly,
these uncertainties do not account for cosmic variance. In the next Section we assess the
errors to be assigned to our best fit parameters by measuring $w(r_p)$
on a series of simulated galaxy catalogs.

A practical integration limit $r_{v0}$ has to be chosen in Eq.~1 in
order to maximize the correlation signal. Indeed, one should avoid
$r_{v0}$ values which are too large since they
would mainly add noise to the
estimate of $w(r_p)$. On the other hand, scales which are too small,
comparable to the redshift uncertainties and to the pairwise
velocity dispersion, should also be avoided since they would not
allow recovery of the entire signal. To search for the best
integration limit $r_{v0}$, we measured $w(r_p)$ and the corresponding
best fit $r_0$ and $\gamma$ values for different $r_{v0}$ values
ranging from 3 to $100\:h^{-1}$ Mpc. Since deviations from a simple
power law are sometimes observed (in particular for
$r_{v0}=20-50\:h^{-1}$ Mpc in GOODS-N), using the best fit correlation
length or clustering amplitude $A=r_0^{\gamma}$ as a measure of the
clustering level is incorrect. To overcome this problem, we chose to
quote the $w(r_p)$ values on a representative scale, as a function of $r_{v0}$.
We adopt $r_p=1$ $h^{-1}$ Mpc as our representative scale, 
since it is well within the considered $r_p$ range, and is a separation 
at which the projected correlation function, $w(1\:h^{-1}$ Mpc$)$, 
is determined with good accuracy.

\begin{figure}
\includegraphics[width=9cm]{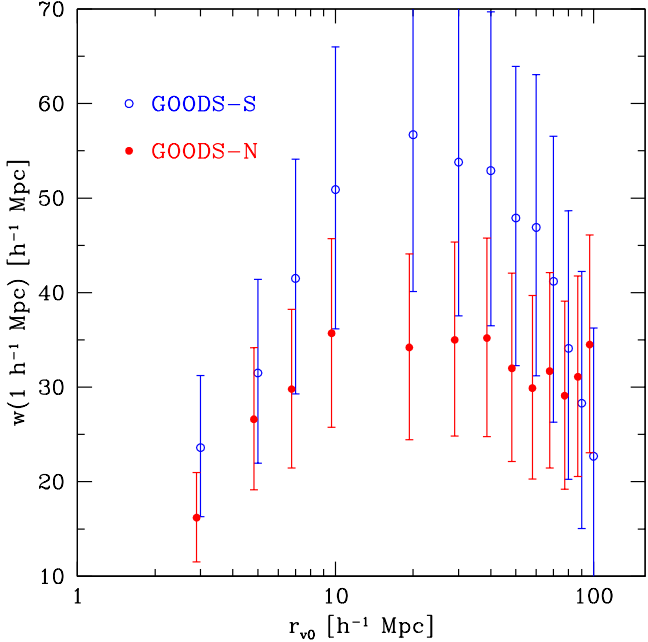}
\caption{Projected correlation function $w(r_p)$ at $r_p=1$
$h^{-1}$ Mpc measured in GOODS-S (open circles) and GOODS-N (filled
circles) as a function of the integration limit $r_{v0}$ (see
Eq.~1). Errorbars take into account cosmic variance (see
Section~4). The turndown at very large scales in GOODS-S is likely
due to sampling noise, in the regime where $r_{v0}$ is much larger
than the size of the redshift peaks (Gilli et al.\ 2003).}
\label{rov}
\end{figure}

In Fig.~\ref{rov} we plot $w$(1 $h^{-1}$ Mpc) as a function of the radial
integration limit $r_{v0}$. We note that the signal amplitude keeps
increasing up to $r_{v0}\sim 10-20\:h^{-1}$ Mpc. For $r_{v0}$ values
greater than $10-20\:h^{-1}$ Mpc, $w$(1 $h^{-1}$ Mpc) does not vary
significantly. In the following, we therefore fix $r_{v0}$ to
$10\:h^{-1}$ Mpc. Such a value for the integration limit is consistent
with what has been widely used in the literature (e.g., Carlberg et
al.\ 2000).

\section{Safety checks and error estimate}

We have checked to see if our method for generating the random sample can bias in
some way the best fit correlation parameters that we measure. In particular,
placing the random sources at the coordinates of the real sources
completely removes the contribution of angular clustering to the total 
clustering signal, which could bias the measured correlation length to lower values. 
We quantify this effect by considering 428 sources within a radius of 4.8
arcmin from the center of GOODS-N, where the spectroscopic
coverage is most complete.  We measured the correlation function in 
two ways:  first, by placing the random sources at the coordinates of the real 
sources, and second, by placing the random sources truly at random within this
area. When using this second method, $r_0$ increases by only 4\%.
Therefore, most of the clustering signal is provided by clustering
along the radial direction, validating the adopted technique.

Another confirmation that this technique is not producing biased
measures comes from tests performed on the mock galaxy catalogs
based on the Millennium Simulation (Springel et al.\ 2005). These
catalogs have been obtained by modeling galaxy formation through
semi-analytic recipes applied to the pure dark matter N-body
simulations of the Millennium run.  Physical processes like gas
cooling, star formation, supernovae and AGN feedback are taken into
account, which are described in detail in Croton et al.\ (2006) and De
Lucia \& Blaizot (2007). Here we considered the most recent work by
Kitzbichler \& White (2007), who built a number of simulated light
cones for deep galaxy surveys over 2~deg$^2$ sky fields. Each cone
contains about $6.5\times10^5$ objects, for which a number of observable
and physical properties like redshift, optical and near-IR magnitude,
and star formation rate are listed. We considered one of these mock
catalogs and applied to the simulated galaxies the same selection
criteria adopted to define our data samples (see details in
Section~\ref{samp}). Here some assumptions have to be made, since
neither the $z_{AB}$ magnitude, not the 24$\mu$m flux density are
directly available for the simulated sources. We used $I_{AB}$ as 
a proxy for $z_{AB}$, assuming the $I-z$ color expected for star forming
galaxies at $z\sim 0.8$ ($(I-z)_{AB}=0.24$, Bruzual \& Charlot
2003). Also, we converted the model star formation rate into IR
luminosity using the relation SFR$=L_{IR}\times
1.72\times10^{-10}\,M_{\odot}$~yr$^{-1}$ and then, at each redshift,
considered only objects above the $L_{IR}$ threshold plotted in
Fig.~\ref{lirz}, which corresponds to the $f_{24}>20\,\mu$Jy threshold
used to define our data samples. The final mock sample contains about
50000 objects, for which we computed the projected correlation
function over the same $r_p$ range used for the GOODS data, first
placing the random control sources at the positions of the Millennium
sources and then placing the random control sources really at random
within the 2~deg$^2$ field. No significant variations are observed
between the projected correlation function computed in the two cases,
suggesting again that the contribution of angular clustering is
negligible.

As shown in Table~1, when the same selection criteria are applied
to the Millennium galaxies, these have on average different redshifts
and luminosities than real mid-IR selected galaxies. We note however
that our main goal is not to select mock galaxies with average
properties identical to the real ones, but investigate any difference
(e.g., in the average $L_{IR}$ or SFR) between the data and the galaxy
formation models once real and mock galaxies have been selected in the
same way. This issue will be addressed in Sections 5 and 6.

\begin{table*}
\begin{center}
\caption{Summary of the best fit clustering parameters. Poissonian 
uncertainties (only) are quoted here to allow comparison between different 
galaxy samples within the same GOODS field (see text). When comparing
the results between the two fields, or when comparing the average
properties of GOODS sources with those of other fields, cosmic
variance uncertainties must also be included (see Table 2).}
\begin{tabular}{lrcccccc}
\hline \hline
Sample& $N^a$& $z$ range& $\bar z^b$& $\bar L_{IR}^c$& $r_0$~~~~~~& $\gamma$& $r_0(\gamma=1.5)$\\
&&&&[$h^{-1}$ Mpc]&&[$h^{-1}$ Mpc]\\
\hline
\multicolumn{7}{c}{GOODS-South}\\
\hline
$f_{24}>20\,\mu$Jy&                     495& 0.1-1.4& 0.74& 4.58& $4.25 \pm 0.12$& $1.51 \pm 0.04$& $4.23 \pm 0.09$\\
$L_{IR}>10^{10}L_{\odot}$&              444& 0.1-1.4& 0.81& 5.51& $4.58 \pm 0.13$& $1.53 \pm 0.04$& $4.52 \pm 0.11$\\
$L_{IR}>10^{11}L_{\odot}$&              161& 0.1-1.4& 1.04& 20.6& $5.22 \pm 0.31$& $1.61 \pm 0.09$& $5.00 \pm 0.29$\\
$L_{IR}\leq 10^{11}L_{\odot}$&          334& 0.1-1.4& 0.67& 2.62& $4.09 \pm 0.15$& $1.54 \pm 0.05$& $4.03 \pm 0.14$\\
$L_{IR}>10^{11}L_{\odot}$&               63& 0.5-1.0& 0.73& 17.2& $6.21 \pm 0.55$& $1.56 \pm 0.14$& $6.12 \pm 0.51$\\
$10^{10}<L_{IR}\leq 10^{11}L_{\odot}$&  177& 0.5-1.0& 0.69& 2.83& $4.18 \pm 0.23$& $1.50 \pm 0.07$& $4.18 \pm 0.17$\\
\hline
\multicolumn{7}{c}{GOODS-North}\\
\hline
$f_{24}>20\,\mu$Jy&                     811& 0.1-1.4& 0.76& 4.26& $3.81 \pm 0.08$& $1.52 \pm 0.03$& $3.77 \pm 0.06$\\
$L_{IR}>10^{10}L_{\odot}$&              734& 0.1-1.4& 0.80& 4.86& $4.03 \pm 0.09$& $1.52 \pm 0.03$& $3.99 \pm 0.07$\\
$L_{IR}>10^{11}L_{\odot}$&              218& 0.1-1.4& 0.95& 20.1& $5.05 \pm 0.27$& $1.55 \pm 0.07$& $4.92 \pm 0.21$\\
$L_{IR}\leq 10^{11}L_{\odot}$&          593& 0.1-1.4& 0.59& 2.78& $3.52 \pm 0.09$& $1.54 \pm 0.04$& $3.46 \pm 0.08$\\
$L_{IR}>10^{11}L_{\odot}$&              111& 0.5-1.0& 0.85& 18.6& $4.66 \pm 0.63$& $1.42 \pm 0.13$& $4.94 \pm 0.40$\\
$10^{10}<L_{IR}\leq 10^{11}L_{\odot}$&  320& 0.5-1.0& 0.75& 3.31& $3.42 \pm 0.11$& $1.67 \pm 0.06$& $3.14 \pm 0.10$\\
\hline
\multicolumn{7}{c}{Millennium$^d$}\\
\hline
$f_{24}>20\,\mu$Jy& 49043& 0.1-1.4& 0.83& 3.6& 2.82& 1.59& 2.52\\
$L_{IR}>10^{10}L_{\odot}$& 44114& 0.1-1.4& 0.87& 4.1& 2.77& 1.58& 2.51\\
$L_{IR}>10^{11}L_{\odot}$& 6423& 0.1-1.4& 1.10& 13.2& 3.31& 1.64& 2.82\\
$L_{IR}\leq 10^{11}L_{\odot}$& 42620& 0.1-1.4& 0.78& 3.0& 2.75& 1.54& 2.63\\
\hline
\end{tabular}
\end{center}

$^a$Number of objects in each sample.\\
$^b$Median redshift.\\
$^c$Median IR luminosity in units of $10^{10}\,L_{\odot}$.\\
$^d$Statistical errors on $r_0$ and $\gamma$ are below 0.01.
\end{table*}

The mock catalogs from the Millennium simulation have also been used
to estimate the global errors on the best fit parameters $r_0$ and
$\gamma$, and to evaluate cosmic variance on the scale of the GOODS fields. 
This has been achieved by extracting from one of the Millennium mock
catalogs samples of galaxies with progressively redder $R-I$ colors
and in the same redshift range as the GOODS galaxies.  The clustering
strength of the mock samples increases with redder $R-I$ color threshold.
We then split the $1.4\times1.4$~deg field over which each sample is distributed
into 40 independent rectangles with the dimensions of a GOODS field
(i.e., $10\times16$ arcmin). For each color sample, we measured the
projected correlation function in each rectangle and computed the {\it
rms} of the $r_0$ and $\gamma$ distributions. After subtracting in
quadrature the (small) term due to Poissonian noise, we are left with
the intrinsic cosmic variance. This procedure allows us to compute the
appropriate variance for sources that are clustered similarly to the 
GOODS galaxies considered.  We found that, on GOODS-sized fields, the
fractional $rms$ of the correlation length increases from 14\% for
sources with $r_0\sim 4\;h^{-1}$ Mpc to 20\% for sources with
$r_0=5.2\;h^{-1}$ Mpc, i.e., for populations as clustered as our total
and LIRGs samples, respectively (see the next Sections). Using the
fractional $rms$ values found with this method, the global errors
related to our samples can be easily estimated once the Poissonian
term is added back in quadrature. When averaging the properties of the
two GOODS fields and presenting the results for the combined GOODS-S
plus GOODS-N sample (see, e.g., Table~2), the variance estimated from
the simulations is divided by a factor $\sqrt 2$.

We note here that the error term due to cosmic variance should only be
considered when comparing the clustering of the same population of
sources across different fields, while it should be ignored when
investigating clustering trends among different source sub-populations
in the same field. Indeed, cosmic variance should increase or decrease
the overall clustering amplitude over a given sky region, without
modifying significantly the relative clustering between different
galaxy subsamples (e.g., sources with different $L_{IR}$), provided
that their redshift distributions are similar, i.e., sources in the
different subsamples are tracing the same large scale structures. 
For this reason, in Table~1 we quote only Poissonian uncertainties,
suitable for comparison between different samples within the same field. 
When comparing the properties of the same population of sources between
GOODS-N and GOODS-S, the cosmic variance term should be included.
When this is done, we find that that the clustering amplitudes measured 
in the two fields are fully compatible with each other (see next Section). 
In Table~2 we quote the clustering parameters averaged between the two 
samples, with uncertainties that include cosmic variance.

The Millennium mock catalogs, in which large source samples can be
selected to minimize statistical noise, were also used to check if
limiting the integration radius $r_{v0}$ to $10\:h^{-1}$ Mpc may
introduce a systematic bias on our clustering measurements. We
selected a population of mock galaxies with $R-I>0.65$, which shows a
clustering level similar to that of our MIPS sources ($r_0\sim
4\;h^{-1}$ Mpc), and measured $w(r_p)$ as a function of the
integration radius $r_{v0}$. We found that for $r_{v0}=30\;h^{-1}$ Mpc
the clustering signal already saturates, and we verified that for
$r_{v0}=10\;h^{-1}$ Mpc the $r_0$ value is biased low by 5\% with
respect to the full, saturated value.  In the Millennium catalogs,
``purely cosmological'' redshifts are also available which are free
from peculiar velocities.  We used these to compute the correlation
function in redshift space $\xi(r)$ for the same mock sample, which
should provide an unbiased measurement of $r_0$. The resulting $r_0$
is in very good agreement with that measured from $w(r_p)$ for
$r_{v0}\geq 30\;h^{-1}$ Mpc and therefore confirms that when using
$r_{v0}=10\;h^{-1}$ Mpc, $r_0$ is biased low by 5\%. We therefore
conclude that the $r_0$ measurements presented in this work could
underestimate the real values by $\sim 5\%$. At any rate, we do not
try to correct for this small systematic bias since it is found to be
well within the uncertainties due to cosmic variance.

Finally, one may wonder if the fitting procedure to $w(r_p)$ adopted
in the previous Section, in which a simple Poisson weighting of the
datapoints is used without considering the effects of cosmic variance,
may bias the best fit parameters $r_0$ and $\gamma$. We verified that,
when attributing to each $w(r_p)$ datapoint the cosmic variance error
as a function of $r_p$ resulting from our simulations, the best fit
parameters $r_0$ and $\gamma$ are essentially unchanged.  In the
GOODS-N field $r_0$ and $\gamma$ change only by $\sim 2\%$. In the
GOODS-S field the change is smaller than 1\%. This is due to the fact
that the datapoints guiding the fits in both procedures are those with
$r_p$ in the range $0.5-4\;h^{-1}$ Mpc, which have both smaller
Poisson errors and cosmic variance. In the following we will therefore
keep using the fitting procedure described in Section~3.

\begin{figure}
\includegraphics[width=9cm]{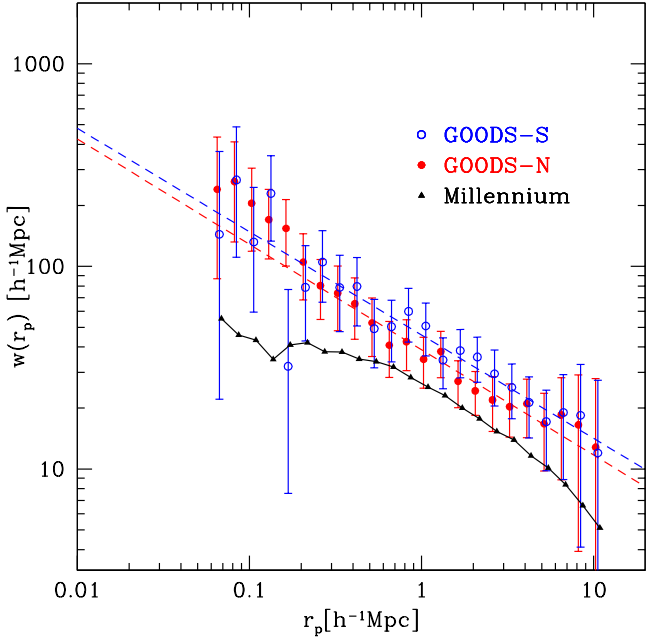}
\caption{Projected correlation function measured for the total
$f_{24}>20\mu$Jy MIPS samples in GOODS-S and GOODS-N (open and filled
circles, respectively) compared with that obtained from the Millennium
simulation on a 2~deg$^2$ field (filled triangles). Errorbars for the
GOODS samples take into account cosmic variance (see Section~4). The
best fit power laws are shown as dashed lines.}
\label{scf}
\end{figure}

\begin{figure}
\includegraphics[width=9cm]{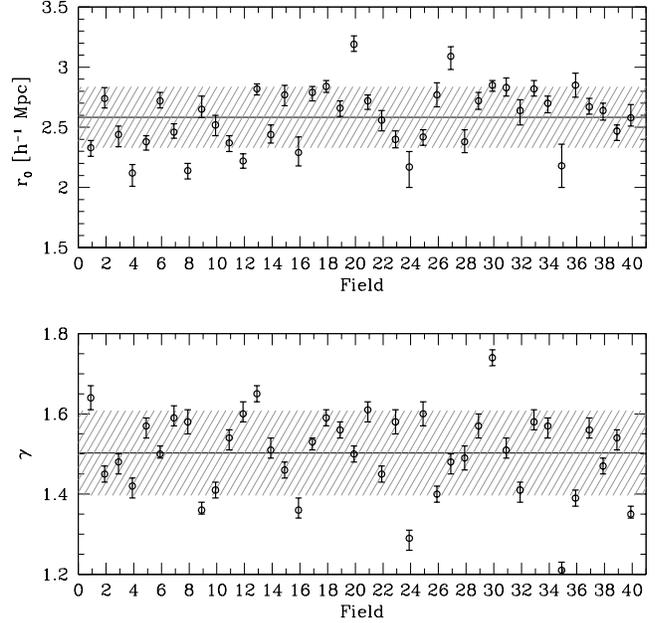}
\caption{Best fit correlation length (upper panel) and slope (lower
panel) measured over 40 mock fields obtained by splitting the 2~deg$^2$
Millennium field into independent rectangles with the dimensions of a
GOODS field. The average $r_0$ and $\gamma$ values (solid lines) and
dispersion (shaded areas) are also shown.}
\label{rodist}
\end{figure}

\section{Results}

Having defined the analysis methods to estimate the galaxy projected
correlation function and the global errors related to it, we are now
in the position to measure the clustering properties of star forming
galaxies in GOODS-S and GOODS-N and to compare them with those expected
for mock galaxies from the Millennium simulation. Also, the
clustering properties of different source subsamples, defined e.g., on
the basis of their IR luminosity, can be readily investigated.

\subsection{Correlation function of the full GOODS-S and GOODS-N samples}

We first measured the projected correlation function for the total
GOODS-N and GOODS-S samples over the projected scale range
$r_p=0.06-10\:h^{-1}$ Mpc. The results are shown in Fig.~\ref{scf}. In
both fields a clear clustering signal is measured, with very high
significance ($>35\sigma$). The best fit parameters ($r_0$, $\gamma$)
are 4.25 $h^{-1}$ Mpc, 1.51 in GOODS-S and 3.81 $h^{-1}$ Mpc, 1.52 in
GOODS-N (see Table~1). The clustering amplitude therefore appears
about 10\% larger in GOODS-S than in GOODS-N, confirming that the
GOODS-S field has more structure than the GOODS-N field, as already
noted from X-ray selected sources (Gilli et al.\ 2005). As shown in
Fig.~\ref{scf}, most of the excess signal in GOODS-S is produced at
projected scales in the range $0.8<r_p<3\:h^{-1}$ Mpc, while at
smaller and larger scales the signals measured in the two fields are
almost identical. A simple check was performed by computing the
projected correlation function in the GOODS-S field after removing
those sources within the two redshift spikes at $z=0.67$ and $z=0.73$,
which showed that most of the excess signal at $0.8<r_p<3\:h^{-1}$ Mpc
is indeed produced by these two structures. At any rate, as it will be
shown later, the difference among the two $r_0$ values is fully
accounted for by cosmic variance.

It should be noted that in the $r_p=0.06-10\:h^{-1}$ Mpc scale range
considered here, the datapoints at the smallest and largest scales are
the least reliable. At small scales, e.g., $r_p<0.1\:h^{-1}$ Mpc, source
pairs at high redshifts ($z>1.2$) have separations on the plane of the
sky comparable to the MIPS angular resolution at $24\mu m$. Therefore
source blending may be an issue. Furthermore, other biases might be
introduced by the different angular selection functions of the many
spectroscopic campaigns from which our catalogs are built. Also, the
transverse size of the GOODS fields (19 arcmin diagonal) becomes
smaller than $r_p\sim 8\:h^{-1}$ Mpc for pairs at $z<0.5$. The
corresponding $w(r_p)$ measurements may thus be distorted with respect
to those at smaller scales because of the different redshift range
sampled. At any rate, the datapoints at the smallest and largest scales
have the largest errorbars and thus do not significantly affect the
overall estimate of the best fit parameters $r_0$ and
$\gamma$. Indeed, when repeating the fits limiting the $r_p$ range to
$0.1-8\:h^{-1}$ Mpc (or even $0.4-8\:h^{-1}$ Mpc), we obtained results
in agreement with the previous ones within the errors. In the
following computations we simply considered datapoints from
$r_p=10\:h^{-1}$ Mpc all the way down to the smallest scale from which
we get signal.

At scales $r_p\lesssim 0.3\:h^{-1}$ Mpc, the correlation function data
points appear to lay above the best fit power law, which may indicate
that the intra-halo clustering term, i.e., the clustering term due to
galaxy pairs within the same dark matter halo, is emerging, as
has recently been seen in very large galaxy samples (e.g., SDSS,
Zehavi et al.\ 2004). However, because of the possible biases in
the $w(r_p)$ datapoints at smaller $r_p$ scales mentioned above, the
observed small-scale excess should be considered with caution. We will
return to this in the Discussion.

The clustering behavior measured for the GOODS samples appears
markedly different from the expectations from the Millennium
simulation. As explained in the previous Section, we computed the
projected correlation function for a sample of about 50000 objects in
a mock galaxy catalog based on the Millennium run after applying the
same selection criteria used for the real data. The projected
correlation function for the mock catalog is also shown in
Fig.~\ref{scf} and the best fit clustering parameters are quoted in
Table~1. Simulated mid-IR selected sources appear much less clustered
than real sources. The overall $w(r_p)$ shape is also very different,
with a flattening below 0.8 $h^{-1}$ Mpc, as opposed to the
steepening observed in GOODS, and a steepening above $r_p\sim 3-4$
$h^{-1}$ Mpc, whereas the GOODS $w(r_p)$ appears to have a constant
slope.\footnote{The subtle differences in the cosmological parameters
adopted in this work with respect to those in the Millennium
simulation ($\Omega_m=0.25$, $\Omega_\Lambda=0.75$, $h=0.73$) are
unlikely to have any significant impact on our results.}

A similar discrepancy between the predictions based on the Millennium
mock catalogs and the real data has also been reported by McCracken et
al.\ (2007), who measured the angular correlation function (ACF) of
$I$-band selected galaxies in the COSMOS field. While at bright
magnitudes the COSMOS and the Millennium ACF are in good agreement, at
fainter magnitudes, $I>22$ mag, Millennium sources are less clustered than
the real COSMOS sources, with an overall correlation function shape
very similar to the one we measured for Millennium. In the same work,
McCracken et al.\ (2007) point out that the observed discrepancy cannot
be accounted for by cosmic variance.

We checked to see if the discrepancy we find can be ascribed to cosmic
variance by dividing the 2~deg$^2$ simulated mock field into 40
non-overlapping rectangles with the same size as that of the GOODS fields
(i.e., $10\times16$ arcmin) and measuring average and dispersion of the
$r_0$ and $\gamma$ distributions over these regions. As shown in
Fig.\ref{rodist}, we found $r_0=2.58$, $\sigma_{r_0}=0.25$ for the
average correlation length and its dispersion, and $\gamma=1.50$,
$\sigma_{\gamma}=0.10$ for the average slope and its dispersion. 
Repeating this exercise on two other independent 2~deg$^2$ mock 
catalogs yielded similar results.

The correlation lengths measured in the GOODS-S and GOODS-N fields then
appear to be about 6 and 5 standard deviations, respectively, larger than 
the value measured from the Millennium catalog. It therefore seems
unlikely that the stronger clustering measured in the GOODS fields be
produced by cosmic variance.  Several possible explanations for
this discrepancy are investigated in the Discussion, as
well as a series of caveats that have to be kept in mind when
comparing models with observations.

It is interesting to note how the average correlation length and slope
measured on these $10\times16$ arcmin mock subsamples are smaller than
those measured for the full 2~deg$^2$ mock catalog and reported in
Table~1. One reason is that at large projected separations, where the
Millennium $w(r_p)$ is steeper, the relative weight of the $w(r_p)$
datapoints is much higher in the full 2~deg$^2$ field than in any
GOODS-sized field, since distant galaxy pairs are much better
sampled. As an example, over the whole $r_p=0.06-10\;h^{-1}$ Mpc range
considered in this work, the number of pairs in a typical GOODS-sized
field is maximum in the range $r_p=3-6\;h^{-1}$ Mpc, while in the full
2~deg$^2$ field it steadily increases towards larger projected
separations. Another reason may be related to the effects of the
integral constraint (Groth \& Peebles 1977), which bias the
measurements of the correlation function on finite size fields. We
estimate that the bias introduced by the integral constraint may
affect the $w(r_p)$ estimates by at most a few percent at the largest
scales probed here (above 5 $h^{-1}$~Mpc).

\begin{table}
\begin{center}
\caption{Combined GOODS-S plus GOODS-N sample.  The uncertainties take into
  account cosmic variance and have been computed as described in
  Section 4.}
\begin{tabular}{lccc}
\hline \hline
Sample& $z$ range& $r_0$[$h^{-1}$ Mpc]& $\gamma$\\
\hline
$f_{24}>20\,\mu$Jy&            0.1-1.4& $4.03 \pm 0.38$& $1.51 \pm 0.08$\\
$L_{IR}>10^{10}L_{\odot}$&     0.1-1.4& $4.31 \pm 0.47$& $1.52 \pm 0.08$\\
$L_{IR}>10^{11}L_{\odot}$&     0.1-1.4& $5.14 \pm 0.76$& $1.58 \pm 0.10$\cr
$L_{IR}\leq 10^{11}L_{\odot}$& 0.1-1.4& $3.81 \pm 0.36$& $1.54 \pm 0.08$\cr
\hline
\end{tabular}
\end{center}
\end{table}


\begin{figure*}
\includegraphics[width=9cm]{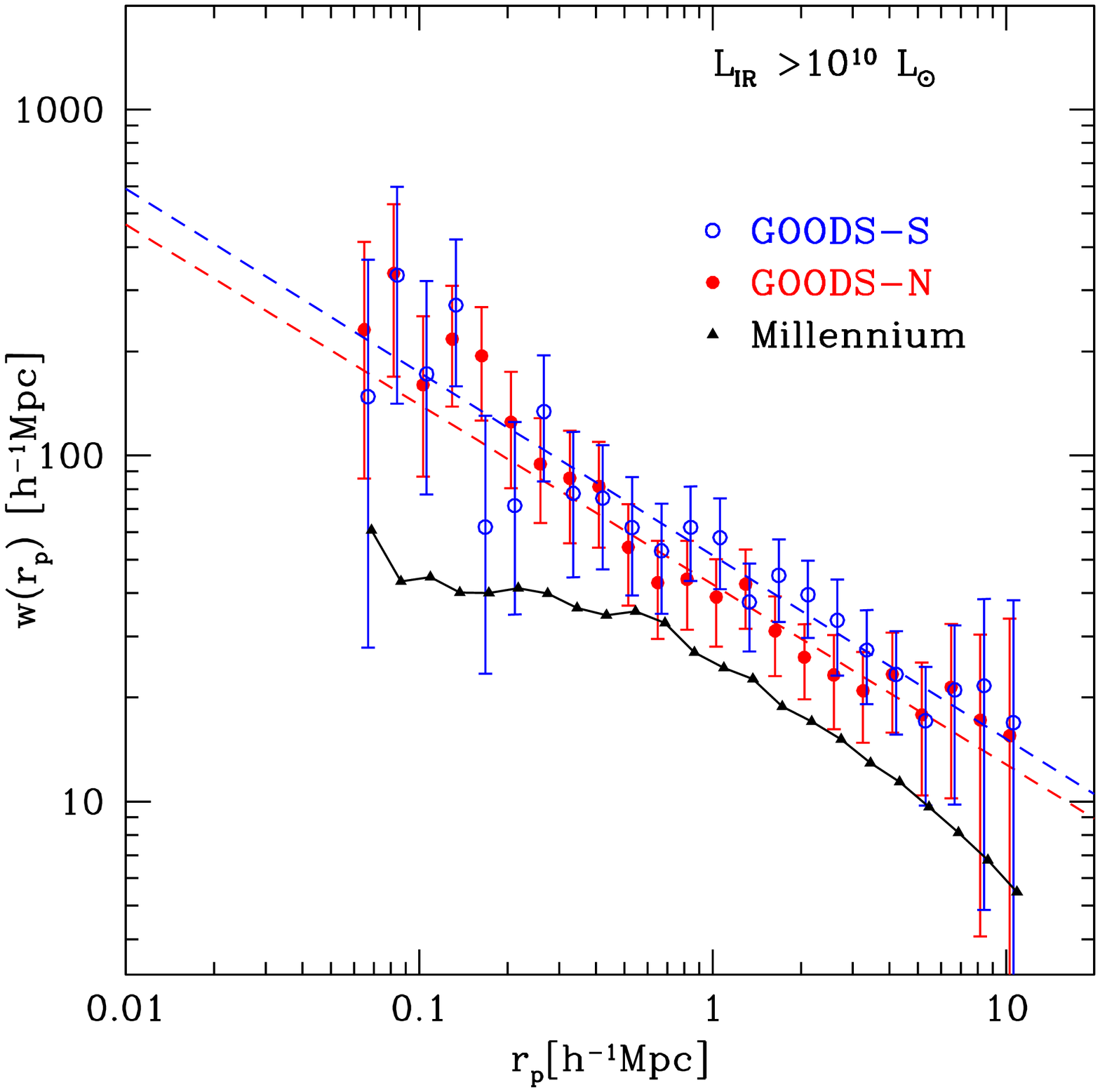}
\includegraphics[width=9cm]{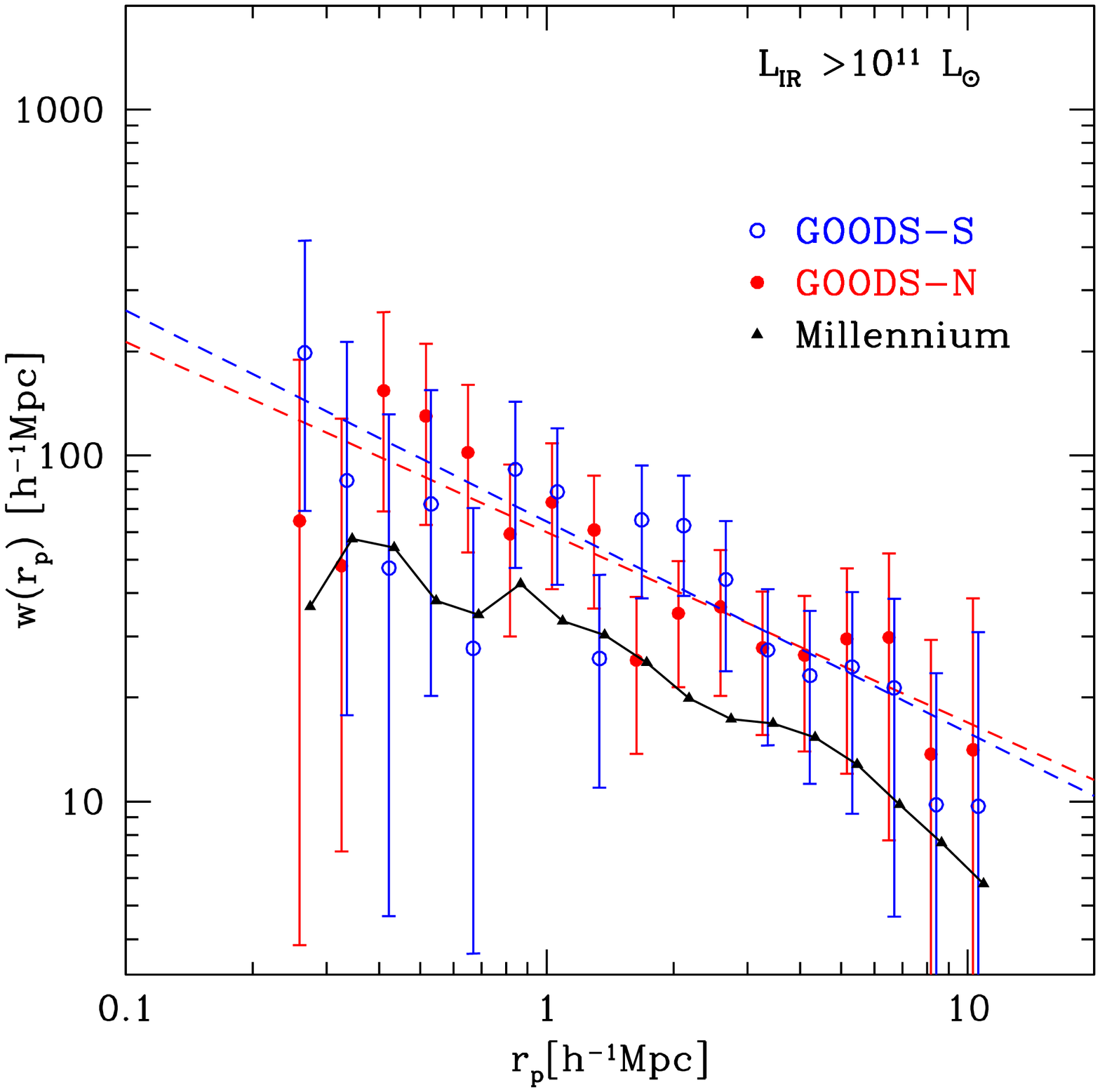}
\caption{$Left$: projected correlation function for sources with
$L_{IR}>10^{10}\,L_{\odot}$ (SFR $>1.7\;M_{\odot}$ yr$^{-1}$) as
measured in GOODS-S, GOODS-N and Millennium simulation (open circles,
filled circles and filled triangles, respectively). Errorbars for
the GOODS samples take into account cosmic variance (see
Section~4). The best fit power laws are shown as dashed lines.
$Right$: as in the $left$ panel but for LIRGs, i.e., objects with
$L_{IR}>10^{11}\,L_{\odot}$ (SFR $>17\;M_{\odot}$ yr$^{-1}$).}
\label{lig}
\end{figure*}

\subsection{Dependence of clustering on IR-luminosity/star formation rate}

Recent observations have shown that, among star-forming galaxies at
any redshift, the star formation rate appears to be correlated with the
galaxy mass (Noeske et al.\  2007; Elbaz et al.\ 2007; Daddi et
al.\ 2007a). This is in agreement with the predictions from
semi-analytic models of structure formation (Finlator et al.\ 2006;
Kitzbichler \& White 2007), though models also predict that this
correlation breaks down for the most massive galaxies. It is therefore
interesting to investigate if and how the clustering of galaxies
depends on the IR luminosity, which is a good proxy for the star
formation rate. We measured the projected correlation function for
sources with $L_{IR}>10^{10}\;L_{\odot}$ and for LIRGs
($L_{IR}>10^{11}\;L_{\odot}$), as shown in Fig.~\ref{lig}. In both
fields we measure an increase of the clustering level with IR luminosity,
with $r_0$ going from $\sim 4\:h^{-1}$ Mpc for the whole
samples to $\sim5\:h^{-1}$ Mpc for the LIRGs (see also Table~1 and 2). A
comparison between the correlation length of the different samples is
shown in Fig.\ref{rolir} for the combined GOODS-S plus GOODS-N
fields. Because of the unavoidable degeneracy between luminosity and
redshift which characterizes any flux limited sample, LIRGs are on
average at higher redshifts than the full IR galaxy
population. However, as reported in Table~1, while the median
luminosity of LIRGs is about a factor of 5 larger than that of the
total sample, their median redshift of $z\sim 1.0$ is not dramatically
higher than that of the total sample, $z=0.75$. The modest difference in
the median redshift for the two samples suggests that luminosity, not
cosmic time, is the main factor contributing to the clustering dependence 
that we observe.  Because the dark matter clustering is smaller at higher
redshift, the difference would be even larger for the implied galaxy bias.
Since $r_0$ for a given galaxy population is expected to increase with
time, i.e., towards lower redshifts (see Section~\ref{evol}), properly
accounting for the redshift differences between subsamples would actually 
strengthen the detection of IR luminosity segregation of clustering.

\begin{figure}[t]
\begin{center}
\includegraphics[width=6cm]{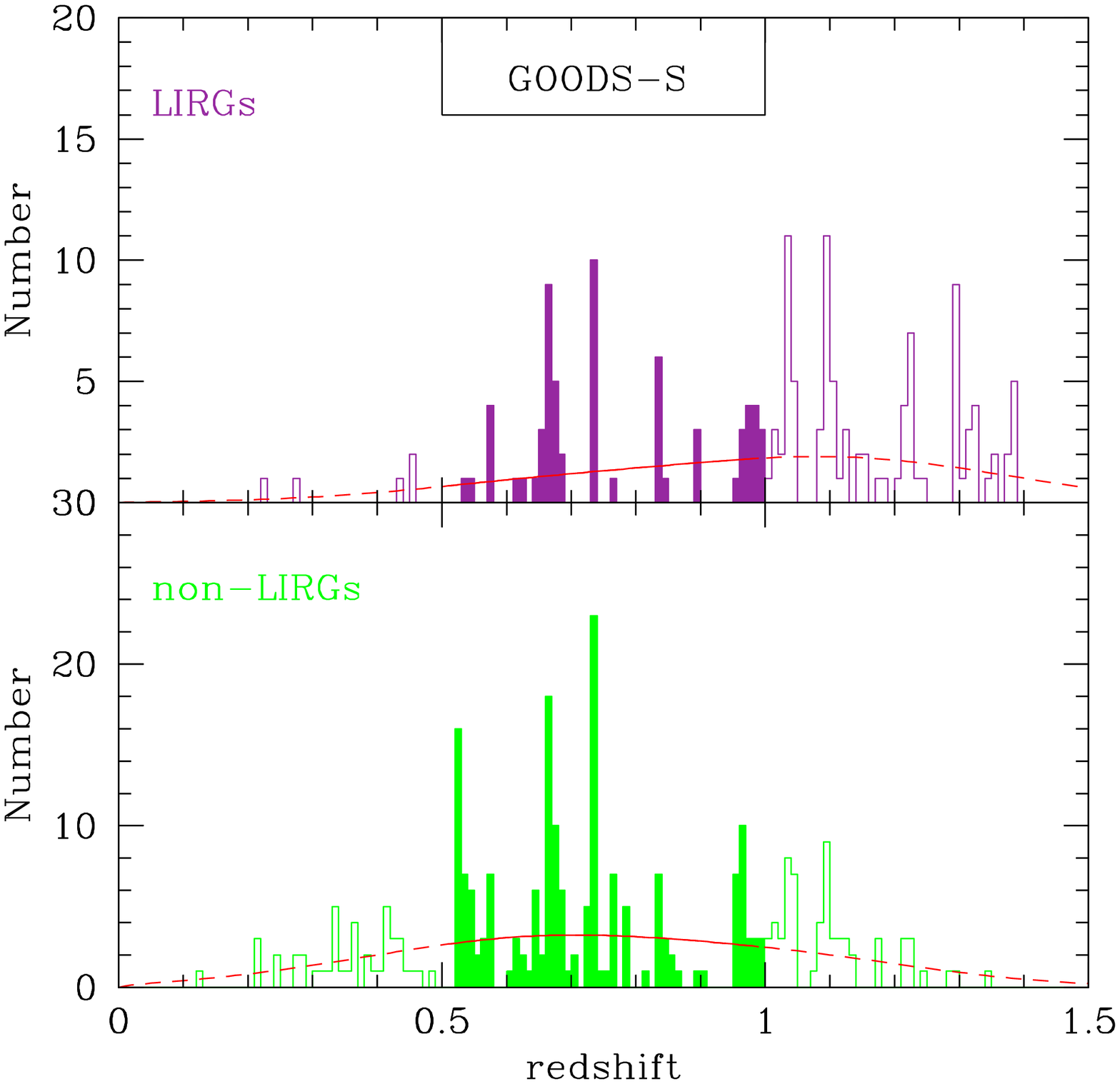}
\includegraphics[width=6cm]{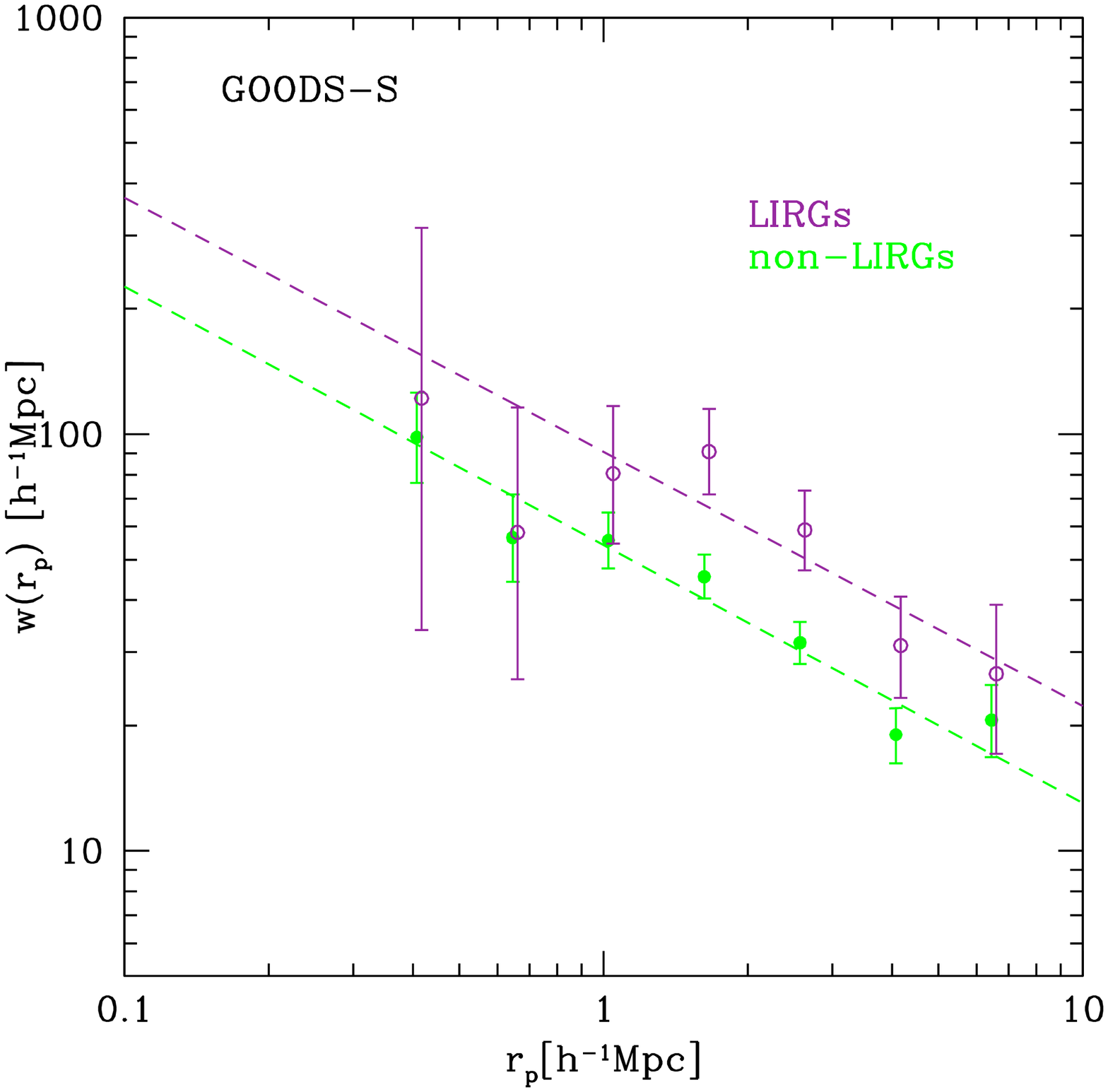}
\caption{{\it Upper panel}: redshift distributions and
selection functions for LIRGs and non-LIRGs in GOODS-S. Sources in the
$z=0.5-1.0$ redshift interval used to compute the projected
correlation function $w(r_p)$ shown in the lower panel have been
shaded. {\it Lower panel}: projected correlation function $w(r_p)$
measured in GOODS-S for LIRGs and non-LIRGs in the redshift interval
$z=0.5-1.0$. Poisson errorbars are used here since the comparison is
performed between samples with similar redshift distributions in the
same field. The best fit power laws are shown as dashed lines.}
\label{noli.cdfs}
\end{center}
\end{figure}

\begin{figure}[t]
\begin{center}
\includegraphics[width=6cm]{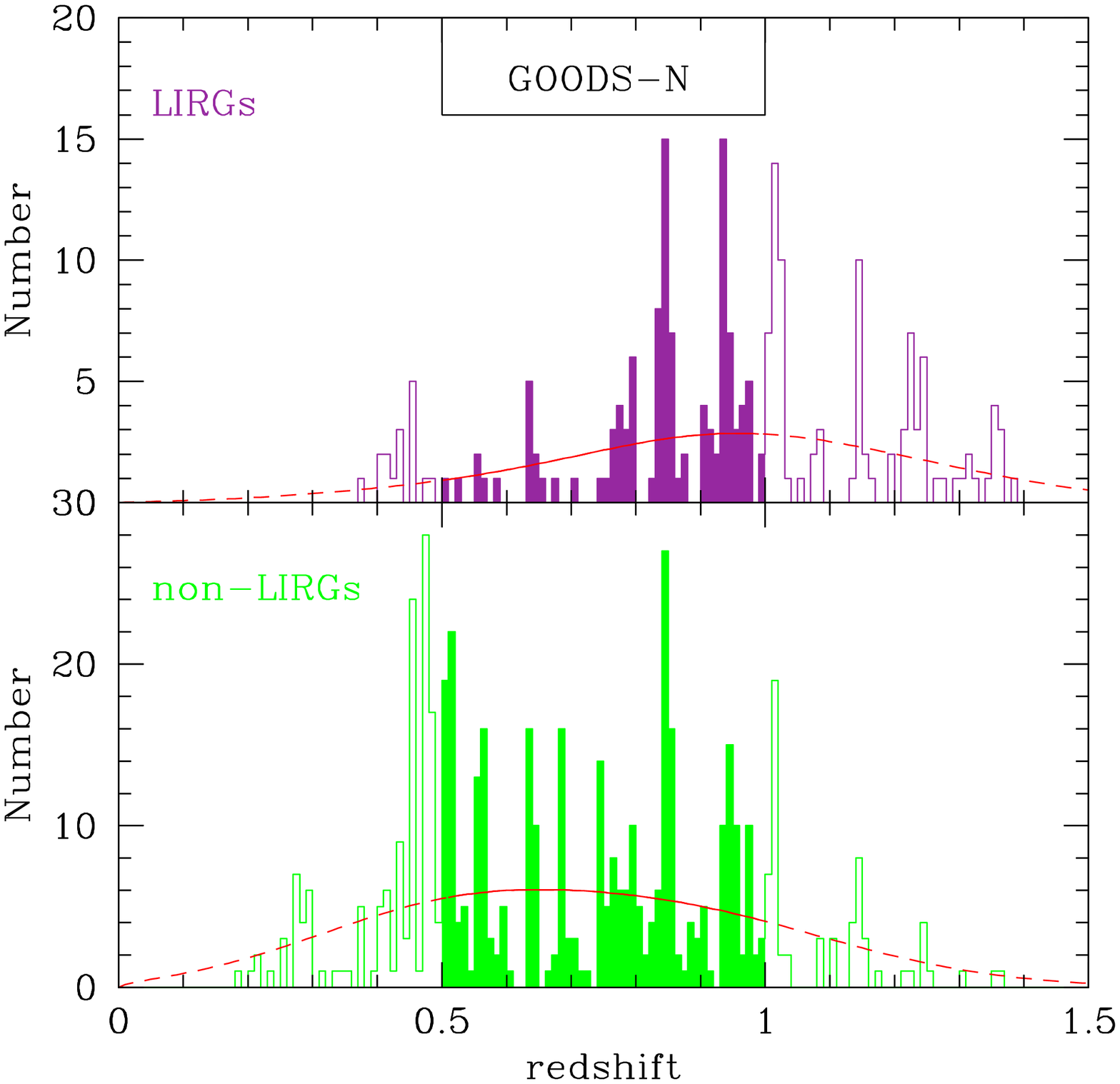}
\includegraphics[width=6cm]{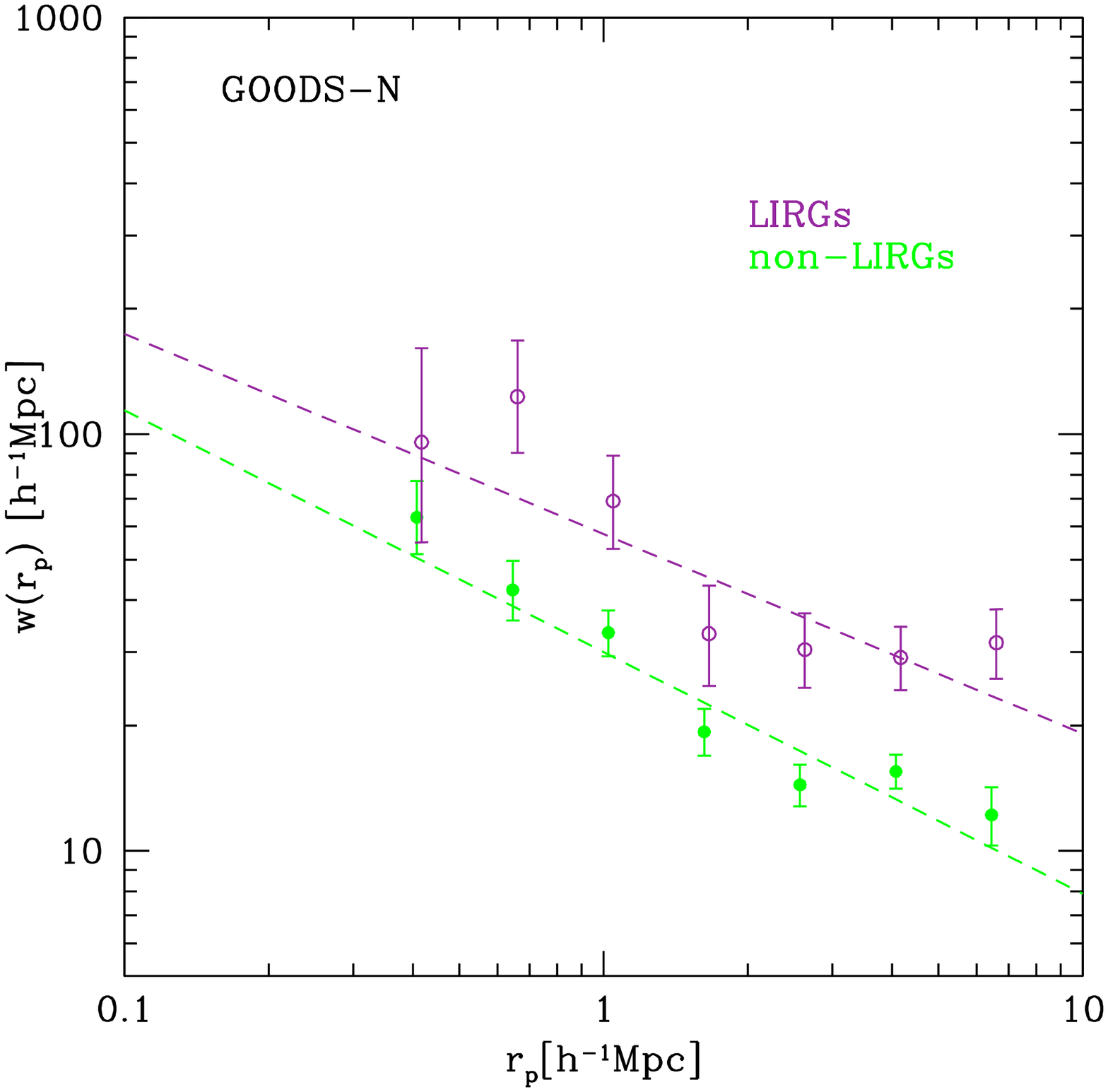}
\caption{Same as in Fig.~\ref{noli.cdfs} but for the GOODS-N field.}
\label{noli.hdfn}
\end{center}
\end{figure}

\begin{figure}
\includegraphics[width=9cm]{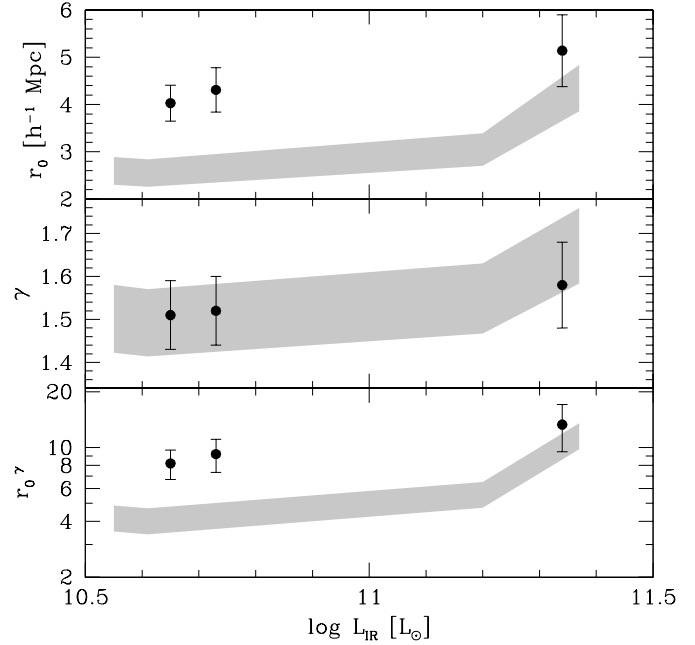}
\caption{From top to bottom panel: best fit correlation length,
slope and amplitude, for the total, $L_{IR}>10^{10}\;L_{\odot}$ and
$L_{IR}>10^{11}\;L_{\odot}$ samples obtained by combining the
GOODS-S and GOODS-N fields (see Table~2). The best fit clustering
parameters are plotted at the sample median $L_{IR}$. The shaded areas
show the average and dispersion of the best fit clustering parameters
measured over 40 mock fields with the dimensions of a GOODS field (see
text for details).}
\label{rolir}
\end{figure}

In order to properly establish the statistical significance of the
trend of clustering versus luminosity, we also considered sources with
$L_{IR}\leq 10^{11}\;L_{\odot}$ (non-LIRGs), which therefore
constitute a source sample disjoint from the LIRGs (see Table~1). The
difference between the clustering correlation length of LIRGs and
non-LIRGs is about $3\sigma$ and $5\sigma$ significant in GOODS-S and
GOODS-N, respectively. As explained in Section 4, only the Poissonian
errorbars quoted in Table~1 have been considered for this estimate.
However, since the redshift distributions of the LIRGs and
non-LIRGs samples are rather different (e.g., the median redshift for
LIRGs is $z \sim 1.0$, while for non-LIRGs it is $z \sim 0.6-0.7$; see
Table~1), this evidence must be investigated further since the two
populations might not be tracing the same large scale structures. We
have therefore restricted our analysis to the redshift range
$z=0.5-1.0$, which allows us to compare LIRGs and non-LIRGs at similar
median redshifts (see Table~1).  Fig.~\ref{noli.cdfs} and
\ref{noli.hdfn} show the redshift distributions and the projected
correlation functions $w(r_p)$ measured for the $z=0.5-1.0$ LIRGs and
non-LIRGs in the GOODS-S and GOODS-N field, respectively. Because of
the limited source statistics, we used larger $r_p$ bins
($\Delta{\rm log}\,r_p$=0.2) than those previously adopted, and limit
our analysis to the $r_p=0.4-8\;h^{-1}$ Mpc range, where the $w(r_p)$
measure is more robust. We found that the significance of stronger
clustering of LIRGs decreases slightly, to $\sim 2-4\sigma$, when performing 
this more appropriate comparison at similar median redshifts. Although the
measured correlation lengths are quite sensitive to the choice of the
redshift bin boundaries because of the spiky nature of the observed
redshift distributions, we note that we systematically measure larger correlation 
lengths for LIRGs than for non-LIRGs, even adopting other redshift intervals.  
We conclude that our data suggest an increase 
of the correlation length with average $L_{IR}$ or SFR, although this 
result needs to be confirmed using larger samples with better statistics.


As in the case of the total sample, we compared the results from the
GOODS fields with those from the Millennium simulation. In
Fig.~\ref{rolir} the $r_0$ values of the samples with
$L_{IR}>10^{10}L_{\odot}$ and $L_{IR}>10^{11}L_{\odot}$ in the
redshift range z=0.1-1.4 for the combined GOODS-S plus GOODS-N sample
(see Table~2) are plotted as a function of the sample median
luminosity and compared with the expectations from mock samples
extracted from Millennium using the same $L_{IR}$ thresholds. Since in
each Millennium sample the median $L_{IR}$ is lower than in the
corresponding GOODS sample (see Table~1) --and this is especially true
for LIRGs-- we also measured $w(r_p)$ for mock sources above
$2\times10^{11}\;L_{\odot}$, which have the same median luminosity of
GOODS LIRGs. Again, we used the 40 GOODS-sized subregions of the
2~deg$^2$ full mock field to obtain the average correlation length and
dispersion for model galaxies selected at different luminosities. This
is shown by the shaded region in Fig.~\ref{rolir}.  Even at
high luminosities, the overall clustering of the data appears stronger
than that predicted by the simulations, although with reduced
significance.

As noted above, among galaxies with $f_{24}>20\mu$Jy, the fraction of
IR luminous objects is lower in the mock catalog than in GOODS. As
an example, the fraction of LIRGs is 13\% in Millenium, as opposed to
the 30\% in GOODS (see Table~1). This is related to the fact that, as
emphasized by Elbaz et al.\ (2007), Millennium galaxies are forming
stars at rates $\approx 3$ times lower than those which are observed at 
$z\sim1$.  We have verified that artificially increasing the SFR of
{\it all} model galaxies (i.e., independent of their positions within 
the simulation) by this amount does not change our conclusions, as it
would imply even smaller correlation lengths all luminosities (as can
already be argued from Fig.~\ref{rolir}).

The AGN removal performed on our sample does not
significantly affect the best fit correlation lengths or slopes.
However, two points are worth noting. First, the fraction of 
AGN candidates is higher among LIRGs (17\%) than in the total samples
(8\%), consistent with what observed for IRAS galaxies in the local Universe,
where a higher fraction of AGN is found in more luminous IR objects
(e.g., Sanders \& Mirabel 1996). Second, a small ($\sim 5-7\%$)
systematic decrease of the correlation lengths is observed when AGN
are removed from the samples, which is consistent with the fact that
AGN in GOODS (which have $r_0=5-10\,h^{-1}$ Mpc, Gilli et al.\ 2005)
are more strongly clustered than is the full IR galaxy population.

\subsection{Implications for the cosmic variance of 24$\mu$m source counts}\label{varsec}

The measured clustering level of star forming galaxies implies that
important field-to-field variations should be observed in the number counts 
of these sources. As discussed in Section~2, we have in fact found that the
surface densities in GOODS-N versus GOODS-S field differ at the 20\%
level, once spectroscopic incompleteness is taken into account.  Given our
direct clustering measurements, we can verify {\it a posteriori} if
this difference may be understood in terms of cosmic variance in the
counts. The expected total variance in the counts can be expressed as:

\begin{equation}
\sigma^2 = N(1+N \times IC)
\end{equation}
\noindent

where $N$ is the average number of galaxies observed and $IC$ is the
integral constraint (see, e.g., Daddi et al.\ 2000 for definitions),
which depends on the angular clustering amplitude $A$ and can be
related to it following Roche et al.\ (1999). We have used the best fit
clustering parameters $r_0$ and $\gamma$, Limber's equation, and the
observed redshift distribution functions (Fig.~\ref{zdist}) to compute
that sources with $f_{24}>20\mu$Jy should have an angular clustering
amplitude of $A(1^o)\sim0.008$. Given the values of the angular
correlation amplitude and slope, and the size of the GOODS fields we
infer an integral constraint of 0.13.  By inserting these values in
Eq.~7 one sees that $N \times IC \gg 1$, i.e., that fluctuations in the number counts
of galaxies with $f_{24}>20\mu$Jy in GOODS-sized fields are dominated
by clustering (i.e., cosmic variance) rather than counting (Poisson)
statistical uncertainties. We expect fluctuations at the level of 35\%
($1\sigma$) in the counts for $f_{24}>20\mu$Jy galaxies in GOODS-sized
fields, fully explaining the observed difference between GOODS-S and
GOODS-N.

\section{Discussion}\label{discussion}

\subsection{Comparison with other galaxy samples at $z\sim 1$}

\begin{figure}
\includegraphics[width=9cm]{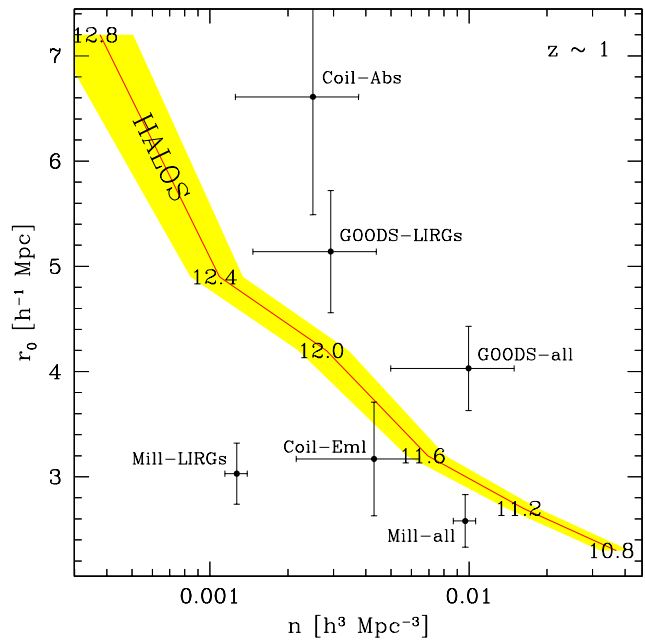}
\caption{Correlation length and space density of GOODS `all'
($f_{24}>20\mu$Jy) and LIRGs galaxy samples considered in this work
are compared to that of other galaxy populations at $z\sim1$, as
labeled.  The trend predicted from the Millennium simulation for dark
matter halos at $z\sim1$ above different mass thresholds is also shown
as a shaded region. More massive halos (log of the threshold mass is
labeled) are less abundant and more clustered than less massive
ones. GOODS IR galaxies and the absorption line galaxies of Coil et
al.\ (2004) appear more abundant than the halos that can host them
(i.e., having the same $r_0$ value), suggesting the presence of more
than one galaxy per halo. As discussed in the text, the corresponding
IR galaxies and LIRGs at $z\sim1$ in the mock galaxy catalogs based on
Millennium appear significantly less clustered than observed in
GOODS. Moreover, Millennium LIRGs are also significantly less
abundant than GOODS LIRGs. Values plotted for Millennium LIRGs and IR
galaxies were derived averaging measurements in 40 GOODS-sized mock
fields.}
\label{nr0}
\end{figure}

Deep redshift surveys such as VVDS and DEEP2 are providing an accurate
census of the galaxy population at $z\sim 1$, measuring in particular
the dependence of galaxy clustering on several parameters such as the
galaxy spectroscopic type, color and luminosity. In both surveys,
galaxies which can be identified as star forming appear to have a
correlation length smaller than that measured for our GOODS 24$\mu$m
selected sample, although the significance of this difference is still
limited. In detail, Coil et al.\ (2004) find $r_0=3.2\pm0.5\,h^{-1}$
Mpc for emission line galaxies in DEEP2 ($\sim1.3\sigma$ lower than
that for the total GOODS $24\mu$m sample), while Meneux et al.\ (2006)
find $r_0=2.5\pm0.4\,h^{-1}$ Mpc for star forming, blue galaxies in
the VVDS ($\sim2.7\sigma$ lower than the total GOODS $24\mu$m
sample). The main difference between the GOODS sample considered here
and those from DEEP2 and VVDS resides in the selection at mid-IR
versus optical wavelengths. The required detection of sources at
24$\mu$m for GOODS (in particular the requirement of
$f_{24}>20\,\mu$Jy) imposes a lower limit to SFR of about 2.5
$M_{\odot}$ yr$^{-1}$ at $z\sim 0.8$ (see Fig.~\ref{lirz}), while
optical selection ($I_{AB}<24$ mag and $R_{AB}<24.1$ mag for VVDS and
DEEP2 galaxies, respectively) does not translate as directly into a
SFR. Indeed, because of older stars and dust extinction, even galaxies
with very similar optical properties could span a very wide range of
star formation rates. We verified that if we impose a cut in SFR or
24$\mu$m flux density on the Millennium mock catalogs, many low-SFR
objects excluded from the sample {\em would} be included if a simple
optical magnitude cut had been used instead (e.g., $z_{AB}<23.5$ mag, the
limit for optical spectroscopy of GOODS sources considered here). In
fact, the median SFR of Millennium mock sources increases by a factor
of $\sim 6$ when the additional mid-IR cut is included.  Therefore, in
optically selected samples, star forming galaxies are expected to have
a lower star formation rate on average than that of our MIPS
sources. The trend discussed in the previous Section, in which $r_0$
is larger for samples selected at increasing $L_{IR}$ (or SFR), is in
line with this interpretation. In connection with the above
considerations, it is interesting to note that the strong clustering
level measured for GOODS LIRGs appears then to be more similar to that
measured for passive galaxies than for moderately star forming
galaxies at $z\sim 1$ (Coil et al.\ 2004, see also
Fig.~\ref{nr0}). Since the amplitude of galaxy clustering is directly
related to the galaxy mass (on average, more massive galaxies reside
in denser, i.e., more clustered, environments), this result is in
agreement with the observed dichotomy for massive galaxies at
$z\lesssim 1.2$, most of which either have already ceased forming
stars, or are doing so at very high rates (Noeske et al.\ 2007; Elbaz
et al.\ 2007).

\subsection{Comparison with predictions of galaxy formation models}

In Section 5 we showed that MIPS detected sources in the GOODS fields
appear to be significantly more clustered than expected from galaxy
formation models based on the Millennium simulation (Kitzbichler \&
White 2007). One may wonder if this discrepancy can be ascribed to
uncertainties in the SFR to $L_{IR}$ conversion, since $L_{IR}$ is the
available (although indirect)
measurement for real data, while SFR is the primary output
for mock galaxies. Under different assumptions on the stellar IMF the
overall uncertainties in the SFR to $L_{IR}$ relation can be
quantified to about 30\%. We verified that a 30\% variation of the
24$\mu$m flux density threshold in the mock catalog does not alter
significantly the Millennium correlation function.

As emphasized by Elbaz et al.\ (2007), at $z\sim 1$ Millennium galaxies
are forming stars at rates about a factor of 3 lower than observed
galaxies. As far as object selection is concerned, artificially
increasing the SFR of model galaxies is equivalent to selecting
galaxies in the mock sample at lower 24$\mu$m flux densities. This
selects many more sources, which are in general less clustered
since the lower tail of the SFR distribution is now being
sampled.  We checked that reducing the limiting $f_{24}$ 
flux density by a factor of 3 produces a lower correlation
function for Millennium sources, thus reinforcing the discrepancy with
the real data. To be fair, it should be noted that simulated galaxies
are free from some of the observational selection effects which affect
real data in our samples and complicate a direct comparison.  For
example, at the faintest flux limits of $f_{24}\approx 20\mu$Jy, where
S/N$\sim5$ for MIPS detections, we might be failing to detect sources
in crowded regions or close to brighter mid-IR targets. We expect this
should be a small effect, but not entirely negligible and in any case
difficult to properly simulate. Also, the 50-65\% spectroscopic
completeness may introduce a bias if sources with measured redshifts
have different clustering properties from sources without redshifts
(i.e., if sources with redshifts are not a random sampling of the full
population). For example, some tendency is detected in both fields for
larger spectroscopic completeness at brighter z-band magnitudes (see
Fig.~1). Therefore the observed discrepancy between the GOODS data and
the mock catalogs from Millennium should be considered by keeping in
mind those caveats. 

At any rate it is interesting to investigate what
could be a likely ingredient that has to be modified within the
semi-analytic models in Millennium to explain the observed
discrepancy. We suggest here that a possible weakness in the models is
the SFR algorithm adopted for the mock galaxies. Indeed, within
simulated dense environments like galaxy clusters and groups, a very
abrupt cut-off of gas-cooling is applied to galaxies as soon as they
become non-central. Therefore, simulated satellite galaxies might be
not forming stars at sufficiently high rates, which would indeed
reduce the correlation length of the star forming simulated population
as well as their number density (see the next Section).

\subsection{The connection with dark matter halos}


While at small scales, comparable to the dimensions of dark matter
halos, the clustering of a given galaxy population is difficult to
predict because of merging and interactions that can trigger a number
of physical processes, at larger scales (e.g., $>1\;h^{-1}$ Mpc), where
galaxy interactions are rare, the galaxy correlation function should
follow that of the hosting dark matter halos.
An interesting consequence is that one can estimate the masses of the
typical halos hosting a given galaxy population by simply comparing
their clustering level (see, e.g., Giavalisco \& Dickinson
2001). Indeed, according to the standard $\Lambda$CDM hierarchical
scenario, dark matter halos of different mass cluster differently,
with the more massive halos being more clustered for any given epoch,
and it is then straightforward to compute the correlation function for
halos above a given mass threshold. It is worth noting that since less
massive halos are more abundant, the correlation function of halos
above a given mass threshold is very similar to the clustering of
halos with mass close to that threshold. Also, it is important to note
that as far as our clustering measurements are concerned (see Section
5), the $w(r_p)$ datapoints at large scales ($r_p>1\;h^{-1}$ Mpc) have
smaller errorbars and guide the power law fit (see Fig.~8). Therefore
the measured $r_0$ and $\gamma$ values are essentially due to the
clustering signal at large scales, where the galaxy correlation
function follows that of the dark matter, allowing a meaningful
comparison with the clustering expected for dark matter halos.

We considered the dark matter halo catalogs available for the
milli-Millennium simulation\footnote{see {\tt
http://www.g-vo.org/Millennium}.}, a reduced version of the Millennium
run which includes 1/512 of the full simulated volume. Halo catalogs
are available at different time steps along the simulation. Here we
considered those at $z\sim1$ (parameter {\tt stepnum=41} in the
simulation). In total there are about 32000 halos with mass above
$10^{10}\;M_{\odot}$ in a cubic volume of 62.5$\;h^{-1}$ Mpc on a
side. We computed the correlation function and the space density of
halos above mass thresholds of log($M/M_{\odot}$)=10.8, 11.2, 11.6,
12.0, 12.4, 12.8. Here we use as halo mass estimator the simulation
parameter {\tt m\_Crit200}, defined as the mass within the radius
where the integrated halo overdensity is 200 times the critical
density of the simulation. The results are shown in Fig.~\ref{nr0},
where it is readily evident that more massive halos are more clustered
and less numerous. The halo region plotted in Fig.~\ref{nr0} takes
into account the fluctuations in the halo space density due to cosmic
variance on volumes equal to the milli-Millennium volume (see
Section~\ref{varsec} and Somerville et al.\ 2004 for a description of
the methods to derive the fluctuations in the source counts from the
clustering parameters).

We computed the space density of sources in our GOODS samples and
compared the $r_0$ and density values of our populations with those of
other galaxy populations at $z\sim1$ and with those of dark matter
halos at $z\sim1$ as computed above. Comparable values for the space
densities of GOODS sources were found when considering the full
$z=0.1-1.4$ redshift range or a restricted redshift interval
($z=0.7-1.2$) around the peak of the selection function. The
comparison is shown in Fig.~\ref{nr0}. Conservative uncertainties of
50\% have been considered in the galaxy space densities, which should
take into account the fluctuations due to cosmic variance as well as
the uncertainties in the volume effectively spanned by the considered
galaxy populations.  By comparing the halo and the galaxy $r_0$
values, one can immediately see that $f_{24}>20\,\mu$Jy star forming
galaxies are hosted by halos with masses $\gtrsim
8 \times 10^{11}\;M_{\odot}$, while LIRGs, which are more clustered, are on
average likely hosted by more massive halos with $M\gtrsim
3 \times 10^{12}\;M_{\odot}$. The population of absorption-line galaxies by
Coil et al.\ (2004) also appears to be hosted by massive halos
($M\gtrsim 5 \times 10^{12}\;M_{\odot}$), while their emission line galaxies
seem to reside in smaller halos with $M\gtrsim
4 \times 10^{11}\;M_{\odot}$. When looking at their space densities,
$f_{24}>20\,\mu$Jy star forming galaxies (and LIRGs) and absorption
line galaxies at $z\sim1$ appear more abundant than halos that can
host them, i.e., there is likely more than one such galaxy per
halo. This is consistent with our measurements of $w(r_p)$. Indeed, as
shown in Fig.~8, the clustering signal is well detected down to very
small scales of $r_p=60\;h^{-1}$ kpc, well within the typical size of
dark matter halos. As an example, the average half-mass radius for
Millennium halos with $M>8 \times 10^{11}\;M_{\odot}$, i.e., those which
likely host GOODS IR galaxies, is about 100 $h^{-1}$ kpc. Therefore,
most of the signal at scales $r_p\lesssim 0.3\;h^{-1}$ Mpc is likely
dominated by galaxies within the same halos (i.e., the so-called
intra-halo term) and a steepening of $w(r_p)$ is indeed consistently
observed at these scales (Fig.~8). A fully consistent analysis of
mid-IR galaxy clustering within the halo occupation number (HOD)
theoretical framework (e.g., Peacock \& Smith 2000; Moustakas \&
Somerville 2002; Kravtsov et al.\ 2004) is however beyond the scope of
this paper.

To conclude this Section we note that Millennium simulated star
forming galaxies and LIRGs at $z\sim1$ are less clustered than
observed in GOODS and that, moreover, observed LIRGs appear
significantly more abundant than those in Millennium (Fig.~12).
This further supports the interpretation that, at $ z\sim1$, many
galaxies within dense environments such as groups or clusters are
forming stars at high rates, in contrast to the star formation history
assumed in the Millennium simulation. The model's scarcity of star forming
galaxies in dense environments, e.g., within the same dark matter halo,
may be also responsible for the observed flattening of the Millennium
correlation function towards small scales (see Fig.~8).

It is not clear yet what is the main driver of star formation in
galaxies at $z\sim1$. On the one hand, a correlation between star
formation rate and galaxy mass is observed (Noeske et al.\ 2007; Elbaz
et al.\ 2007). On the other hand, as found in this work, higher star
formation rates are hosted by galaxies in denser environments. These
two results are perfectly consistent one another (and with the
conclusions of Elbaz et al.\ 2007 and Cooper et al.\ 2007), since more
massive galaxies are indeed located in dense environments, but it is
hard to establish what is the ultimate driver for the star formation
increase: is it the galaxy mass or the environment? In other words, is
the star formation rate in each galaxy simply linked to the gas mass
and triggered at a given time along the galaxy life almost
independently of the environment or, instead, are environmental
effects necessary to produce gas instabilities and trigger star
formation? Solving these issues is beyond the scope of this paper. It
will require much larger samples of star forming galaxies with
spectroscopic redshifts, with which one will be able to study
clustering of galaxies versus their star formation rates in narrow
mass bins.

\begin{figure}
\includegraphics[width=9cm]{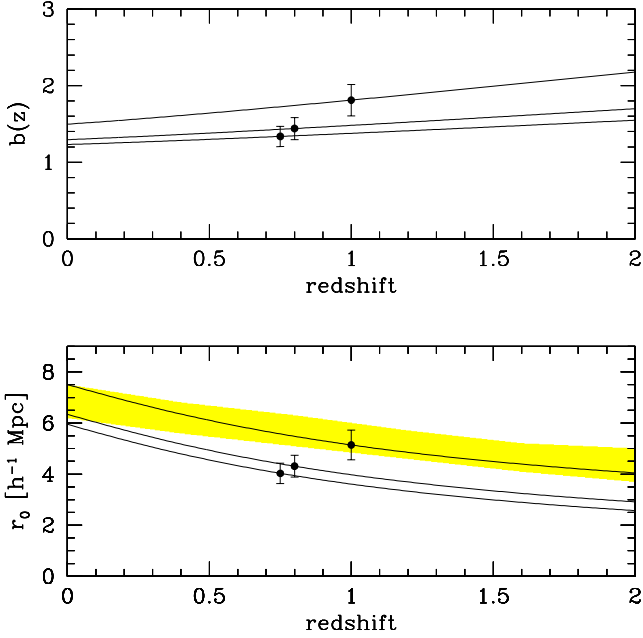}
\caption{Bias (upper panel) and correlation length (lower panel) for
the total, $L_{IR}>10^{10}\,L_{\odot}$ and $L_{IR}>10^{11}\,L_{\odot}$
combined GOODS samples quoted in Table~2, compared with evolutionary
tracks computed according to a {\it conserving scenario} (solid lines,
see text for details). The shaded area shows the $r_0$ evolution of
$z\sim 2$ star forming galaxies as computed by Adelberger et
al.\ (2005).}
\label{broz}
\end{figure}

\subsection{Descendants and progenitors  of $z\sim0.5$--1 star forming galaxies}\label{evol}
Under simple assumptions, the spatial clustering of an extragalactic
source population measured at a given epoch can be used to estimate
the typical dark matter halos in which these objects reside, and then
to estimate their past and future history by following the halo
evolution in the cosmological density field. A useful quantity for
such analyses is the bias factor, defined as
$b^2(r,z,M)=\xi_g(r,z,M)/\xi_m(r,z)$, where $\xi_g(r,z,M)$ and
$\xi_m(r,z)$ are the correlation function of the considered galaxy
population and that of dark matter, respectively. In general the bias
parameter can be a function of scale $r$, redshift $z$, and object
mass $M$. For simplicity we adopt the following definition here:
\begin{equation}
b^2(z)=\xi_g(8,z)/\xi_m(8,z)
\label{bb}
\end{equation}
in which $\xi_g(8,z)$ and $\xi_m(8,z)$, are the galaxy and dark matter
correlation function evaluated at 8 $h^{-1}$ Mpc, respectively. The
galaxy correlation function has been measured directly in this work,
while the dark matter correlation function can be estimated using the
following relation (e.g., Peebles 1980):
\begin{equation}
\xi_m(8,z)=\sigma_8^2(z)/J_2
\end{equation}
where $J_2=72/[(3-\gamma)(4-\gamma)(6-\gamma)2^\gamma]$ and
$\sigma_8^2(z)$ is the dark matter mass variance in spheres of 8
$h^{-1}$ Mpc comoving radius, which evolves as
$\sigma_8(z)=\sigma_8(0)D(z)$.
$D(z)$ is the linear growth factor of perturbations, while
$\sigma_8=\sigma_8(0)$ is the $rms$ dark matter fluctuation at present
time, which we fix to $\sigma_8=0.8$ in agreement with the recent
results from WMAP3 (Spergel et al.\ 2007). While in an Einstein - De
Sitter cosmology the linear growth of perturbations is simply
described by $D_{EdS}(z)=(1+z)^{-1}$, in a $\Lambda$-dominated
cosmology the growth of perturbations is slower. We consider here the
so-called growth suppression factor $g(z)=D(z)/D_{EdS}(z)$ as
approximated analytically by Carroll, Press \& Turner (1992).


The above relations allow us to estimate the bias of the galaxy
population at its median redshift. One can further assume that the
spatial distribution of the observed galaxy population simply evolves
with time under the gravitational pull of growing dark matter
structures. This scenario, in which galaxy merging is considered
negligible, is often called the {\it galaxy conserving model} and in
this case the bias evolution can be approximated by
\begin{equation}
b(z)=1+[b(0)-1]/D(z)
\end{equation}
where $b(0)$ is the population bias at $z=0$ (Nusser \& Davis 1994, Fry
1996, Moscardini et al. 1998).
  
Once $b(z)$ is determined, the evolution of $\xi_g(8,z)$ and hence of
$r_0(z)$ can be obtained by inverting Eq.~\ref{bb}. The best fit
$\gamma\sim 1.5$ values found in this work are assumed in the above
relations. In Fig.~\ref{broz} we show the evolution of $b(z)$
and $r_0(z)$ for the combined GOODS samples reported in Table~2. Star
forming ($f_{24}>20\;\mu$Jy) objects at $z\sim0.7$ are expected to
have $r_0\sim6-7\:h^{-1}$ Mpc at a redshift of 0.1. Since local early
type galaxies \footnote{In the $R$ band, the
characteristic luminosity of $z\sim 0$ early type galaxies $L_*$ is
$M^*_R=-21.5$ (Baldry et al.\ 2004).} with $L<L_*$ have been observed to be
clustered that strongly in the SDSS and 2dFGRS (Zehavi et al.\ 2002;
Madgwick et al.\ 2003), at least part of them could descend from
$z\sim0.7$ star forming objects. Similarly, some of the brighter
($L\sim L_*$) ellipticals in the local Universe, for which
$r_0\sim8\,h^{-1}$ Mpc has been measured (Guzzo et al.\ 1997, Budavari
et al.\ 2002) could descend from $z\sim1$ LIRGs
($L_{IR}>10^{11}\,L_{\odot}$), which are expected to evolve into a
population with $r_0\sim7-8\:h^{-1}$ Mpc by $z=0$. This would be
consistent with the recent findings by Cimatti, Daddi \& Renzini
(2006), who observe a lower number density of $L\lesssim L_*$ early
type galaxies at $z\sim 0.8$ than at $z=0$, suggesting that at least
part of local ellipticals have formed since $z\sim1$.

The slope of the correlation function for local ellipticals is
generally found to be steeper than that observed for GOODS IR
galaxies. Slopes of $\gamma\sim 1.9-2$ have indeed been measured for
local ellipticals (Guzzo et al.\ 1997, Zehavi et al.\ 2002, Madgwick et
al.\ 2003), as opposed to $\gamma\sim1.5-1.6$ for GOODS star forming
galaxies measured in this work. While an average steepening of the
matter correlation function and of the overall galaxy population is
expected towards lower redshifts (see, e.g., Kauffman et al.\ 1999,
Moustakas \& Somerville 2002) since the clustering level progressively
increases at smaller scales, the clustering evolution in the proposed
{\it galaxy conserving scenario} above is computed by assuming a fixed
($\gamma=1.5$) slope. Also it has to be kept in mind that the {\it
galaxy conserving scenario} is an ideal, rather extreme,
representation of galaxy evolution, since it, by definition, neglects
galaxy merging. It is therefore somewhat misleading to determine the
descendants of a high redshift galaxy population simply based on the
$r_0$ comparison without considering the slope. A $z\sim1$ star
forming galaxy does not evolve automatically into a $z=0$ elliptical
and perhaps subsamples of the local spiral galaxy population may have
the clustering properties expected for the descendants of $z\sim1$
star forming galaxies. In an SDSS-based paper, Budavari et al.\ (2002)
have analyzed the clustering properties of $z\sim 0.2$ galaxies with
different spectral energy distributions (SEDs) corresponding to those
of galaxies with different morphological types. They found that bright
($-23<M_R<-21$) galaxies with SEDs corresponding to the morphological
type Scd have a correlation length of $r_0=6.75\,h^{-1}$ Mpc, similar to
those of ellipticals at the same redshift, but with a shallower slope
$\gamma\sim1.7$. We suggest that part of the GOODS LIRGs population
may then evolve into bright, massive spirals. By adding the space
densities of local ellipticals and bright spirals one further sees
that this is similar to what is measured for $z\sim1$ star forming
galaxies.

Recently, Adelberger et al.\ (2005) measured the clustering of star
forming galaxies at $z\sim1.5$--2 (BM and BX samples) and at $z=3$
(LBGs, see also Giavalisco \& Dickinson 2001). By comparing the galaxy
correlation function with that of dark matter halos in the
$\Lambda$CDM-GIF simulation (Kauffmann et al.\ 1999), Adelberger et
al.\ (2005) found that UV selected galaxies at $z\sim 2$ are hosted by
halos with masses around $10^{12}\,M_{\odot}$. Furthermore, by
following the evolution of these halos in catalogs computed at
subsequent time steps in the simulation, they were then able to infer
the correlation length of the descendants of the $z\sim 2$ galaxy
population. At $z\lesssim1$ they find that the only galaxy population
with clustering strong enough to be consistent with that of the expected
descendants of UV selected galaxies are red absorption line dominated
galaxies from Coil et al.\ (2004). In Fig.~\ref{broz} the expected
evolution of $z\sim2$ starburst galaxies as computed by Adelberger et
al.\ (2005) is also shown. The clustering length of LIRGs at $z\sim1$
is large enough to be consistent with the one predicted for the
descendants of UV selected galaxies. Moreover, the correlation slopes
of the two populations are similar ($\gamma\sim1.5-1.6$). The
average SFR of UV-selected galaxies is also of the same order of that
of LIRGs (35 $M_{\odot}$ yr$^{-1}$ on average.) It is therefore
possible that LIRGs at $z\sim0.5$--1, in addition to passive galaxies,
may be the direct descendants of UV-selected galaxies. This would
imply, in turn, that star formation in these galaxies is sustained,
either continuously or intermittently, over cosmological timescales of
a few Gyrs and suggests they assemble stellar masses up to $\sim
10^{11}\;M_{\odot}$ from $z\sim3$ to $z\sim1$. Our conclusions on the
$z\sim1 $ descendants of high redshift star forming galaxies add to
those reached by Adelberger et al.\ (2005), who, based on the
comparison with the correlation lengths measured in the DEEP2 surveys,
identify passive absorption line galaxies at $z\sim 1$ as the
descendants of their LBG population. DEEP2 star forming objects were
on the contrary ruled out based on their small correlation length. As
explained in the previous Section, the low correlation length of
emission line (star forming) galaxies in the DEEP2 survey can be
ascribed to a SFR on average lower than that measured for our
LIRGs. Our results suggest that star formation is intense in a
significant fraction of massive objects at $z\sim 1$ and that the
descendants of high redshift star-forming galaxies have not
necessarily stopped forming stars at $z\sim0.5$--1. If we consider
that LIRGs and passive galaxies at $z\sim 1$ have similar space
densities ($\sim 2.5-3\times\;10^{-3}$~Mpc$^{-3}$, Fig.~\ref{nr0}),
and that their combined density is of the order of the LBG space
density ($\sim 4-6\times10^{-3}$Mpc$^{-3}$), then we can conclude that a
significant fraction of $z\sim 2$ star forming galaxies might still be
forming stars at $z\sim 1$.


\section{Summary and conclusions}

We present the first measurements of the spatial clustering of star
forming galaxies at $z\sim 1$ selected at 24$\mu$m by {\it
Spitzer}/MIPS in the GOODS-S and GOODS-N fields. The correlation
length for the total combined sample has been found to be
$4.0\pm0.4\:h^{-1}$ Mpc, the $r_0$ value in GOODS-S being $\sim10\%$
larger than in GOODS-N. We estimate the uncertainties in our
measurements using mock catalogs extracted from the Millennium
simulation, which show that the GOODS-S and GOODS-N measurements are
fully consistent with the expected cosmic variance on these 160
arcmin$^2$ fields. We find indications for an increase of the
correlation length with $L_{IR}$ (or SFR), with LIRGs having
$r_0\sim5.1\pm0.8\:h^{-1}$ Mpc. The measured correlation length in the
GOODS mid-IR selected samples appears larger than that measured in
optical samples of star forming galaxies at $z\sim1$ such as those in
the DEEP2 or the VVDS surveys. Although the significance of this
result is still limited ($1-3\sigma$), it might be interpreted as
evidence that the average star formation rate in optically selected
samples of emission line galaxies is lower than that of our samples,
which, by selection, have larger IR luminosity. This is in agreement
with the observed relation between IR luminosity and clustering
strength, which, in turn, suggests that at $z\sim 1$ more intense star
formation is hosted by more massive (i.e., more clustered) systems.

The measured correlation length is significantly larger than that
expected from the Millennium simulations, once the selection criteria 
adopted to define the real data samples are applied to the mock 
samples. This suggests that star formation is, on average, occurring 
in dark matter halos that are more massive than those predicted by
the galaxy formation model implemented in the Millennium simulation
by Croton et al.\ (2006).  By comparing the clustering of GOODS star
forming galaxies with that of Millennium dark matter halos, we find
that more luminous galaxies are hosted by progressively more massive
halos, with LIRGs residing in halos with $M\gtrsim
3 \times 10^{12}\,M_{\odot}$. Since the measured LIRG space density is
higher than that of the hosting halos, each halo appears to contain on
average more than one LIRG. This is also supported by the steepening
of the correlation function observed towards smaller scales, which is
usually interpreted as due to galaxy pairs within the same dark matter
halo (intra halo clustering).

Based on a galaxy conserving scenario, in which it is assumed that
galaxies observed at a given redshift evolve without merging, simply
pulled by the surrounding density field, we trace the time evolution
of the bias parameter and of the correlation length of $z\sim 1$
star forming galaxies. By comparing the evolved correlation lengths
with those of local and high-redshift galaxy samples, we infer the
likely descendant and progenitors of our $z\sim 1 $ sample.  We find
that objects in our sample may evolve into $L<L_*$ ellipticals or
bright spirals by $z=0$, with LIRGs evolving into bright $L\sim L_*$
objects. Similarly, LIRGs, together with passive absorption line
galaxies at $z\sim 1$, may be identified as the descendants of
UV-selected star forming galaxies at $z\sim 2$.

\acknowledgements

We wish to thank the referee for comments which improved the paper
significantly. We acknowledge G.\ Zamorani, L.\ Pozzetti, F.\ Pozzi,
C.\ Gruppioni, L.\ Moscardini, E.\ Branchini and M.\ Magliocchetti for
useful discussions. We are also grateful to E.\ MacDonald and H.\ Spinrad
for their extensive efforts obtaining some of the redshift measurements 
used in this work. R.G. acknowledges financial support from the Italian
Space Agency (under the contract ASI-INAF I/023/05/0) and from the
grant PRIN-MUR 2006-02-5203.  The work of D.S. was carried out at Jet
Propulsion Laboratory, California Institute of Technology, under a
contract with NASA.

\end{document}